\documentclass{elsart}
\journal{EPJ C}

\usepackage{epsfig}

\begin{document}
\begin{frontmatter}

\title{Axes determination for segmented true-coaxial HPGe detectors}
\author[a]{I.~Abt},
\author[a]{A.~Caldwell},
\author[a,b]{J.~Liu},
\author[a]{B.~Majorovits},
\author[c]{P.~Petrov},
\author[a]{O.~Volynets\corauthref{cor}\ead{volynets@mppmu.mpg.de}}

\address[a]{Max-Planck-Institut f\"ur Physik, M\"unchen, Germany}
\address[b]{now: Institute for the Physics and Mathematics of the Universe, Tokyo University, Tokyo, Japan}
\address[c]{Durham University, Department of Physics, United Kingdom}

\corauth[cor]{Max-Planck-Institut f\"ur Physik, F\"ohringer Ring 6, 
              80805 M\"unchen, Germany,
              Tel.: +49-(0)89-32354-415, Fax: +49-(0)89-32354-528}

\date{\today} % Mar 26 2012

\begin{abstract}
A fast method to determine the crystallographic axes of segmented
true-coaxial high-purity germanium detectors is
presented. 
It is based on the analysis of 
segment-occupancy patterns obtained
by irradiation with radioactive sources.
The measured patterns are compared 
to predictions for different axes orientations.
The predictions require a simulation of the
trajectories of the charge carriers
taking the transverse anisotropy of their drift 
into account.
\end{abstract} %end of abstract
\end{frontmatter}

\maketitle
%

% --------------------------------------------------------
% body
% --------------------------------------------------------
\section{Introduction}
\label{section:introduction}
High purity germanium detectors, HPGeDs, are used in a wide variety 
of applications in particle and nuclear physics 
\cite{Kno99,eberth,agata+greta,GERDA}. 
Segmented detectors can be used to disentangle event topologies
\cite{pos_Agata,photon,neutron}. The detectors considered here
are segmented cylindrical true-coaxial detectors. 
The segmentation in azimuth angle \footnote{A cylindrical coordinate
system is used with the origin at the center of the detector.} $\phi$
is the decisive feature for the proposed method.

The orientation of the crystallographic
axes of a germanium detector is important 
for many analyses where pulse shapes are used.
The difference between the charge-carrier mobilities
along the
crystallographic axes  $\langle 110 \rangle$ and 
$\langle 100 \rangle$  is significant.
It is called longitudinal anisotropy.
Experimental data on hole mobility exist for the $\langle 111 \rangle$ and
$\langle 100 \rangle$ axes \cite{bart}. Relevant here is the plane containing the
$\langle 110 \rangle$  and $\langle 100 \rangle$ axes.
Appropriate calculations \cite{pss} imply 
that the difference of the velocities of the holes along the
$\langle 110 \rangle$ and $\langle 100 \rangle$ axes is about 10\,\% for
realistic values of the electric field. 
The different charge carrier velocities result in different rise times
of pulses originating at equal distances from the electrode.
The differences in rise times between pulses along the 
$\langle 110 \rangle$ and $\langle 100\rangle$ is typically around 10\% \cite{pss,char}
depending on the relative location of the charge deposits to the
crystal axes.
In addition, the paths of charge carriers not drifting along
the crystallographic axes are bent; 
this is called transverse anisotropy.
The anisotropies have to be taken into 
account in so called pulse shape
analyses \cite{psam,psa} and whenever 
simulated pulses \cite{pss} are compared
to data.

The crystallographic  $\langle 001 \rangle$-axis
of a cylindrical germanium detector is 
usually aligned with the $z$ axis.
The position of the  $\langle 110 \rangle$ axis, 
$\phi_{\langle 110 \rangle}$, is, however, {\it a priori}
not known after the detector is processed.
Therefore the orientation of the $\langle 110 \rangle$ axis with 
respect to the segment boundaries
has to be determined during the characterization of the detector.

This paper presents a comparison of two methods to determine the axes.
The first method is based on scanning the detector.
It is widely used, well proven  and based on data only.
However, it requires that the
detector is mounted in a special test setup with a movable source
and it is time consuming.
The second method involves the comparison to
Monte Carlo simulation. Its advantages are that it is fast and can
be performed in any configuration, even if the detector is part of a complex
detector system.
The determination of the axes is required to have an accuracy of
about 5$^{\circ}$. A simple application using the
orientation of the axes is the calculation of the
probability that a neutrinoless double beta decay event
actually produces a signal in two segments of a detector, even though
it is a so called single-site event. This requires that the knowledge
about the axes orientation is sufficient to calculate whether
a drifting charge cloud gets separated across segment boundaries.
The charge cloud representing a neutrinoless
double beta decay event has a diameter of about 2\,mm \cite{r90},
and thus covers about 5$^{\circ}$ at a radius of 25\,mm in a
cylindrical detector.

\section{Detector and test environment}
\label{section:test_env}

\begin{figure}[htbp]
  \begin{minipage}{0.45\textwidth}
    \centering
    \includegraphics[width=\textwidth]{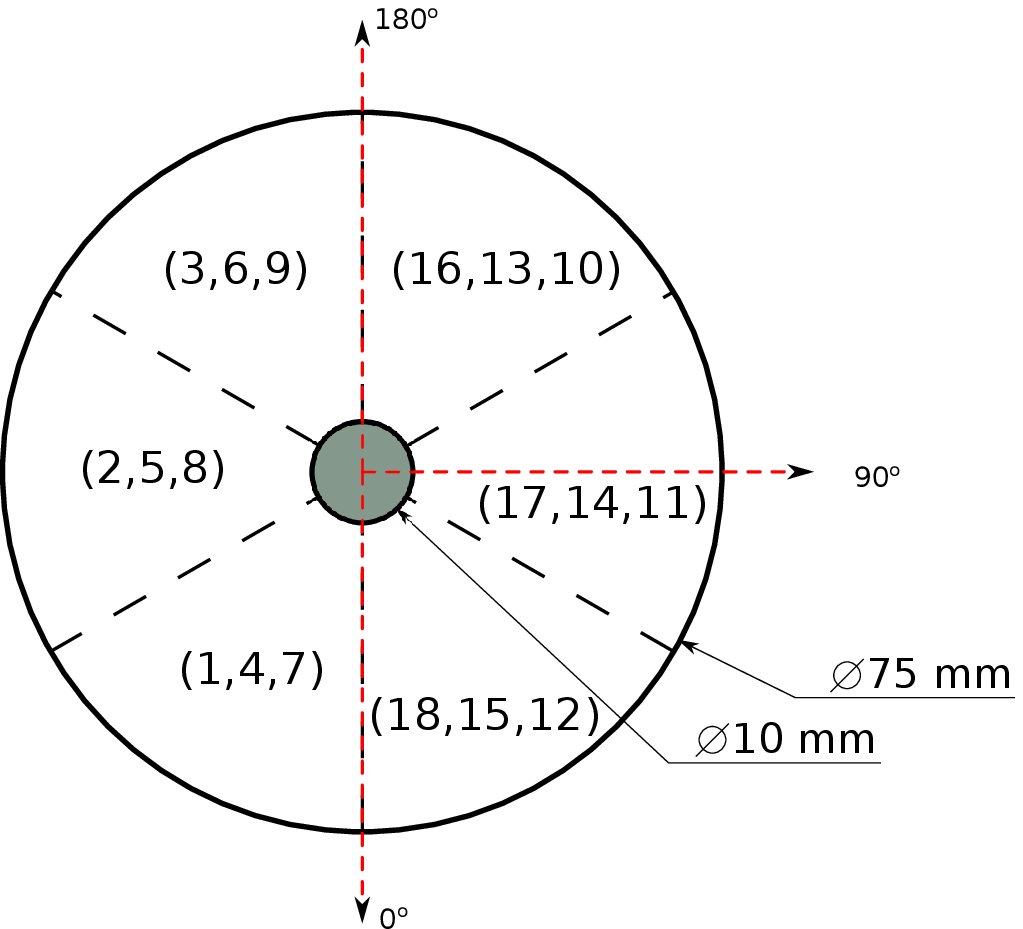}
  \end{minipage}
  \begin{minipage}{0.52\textwidth}
    \centering 
    \includegraphics[width=\textwidth]{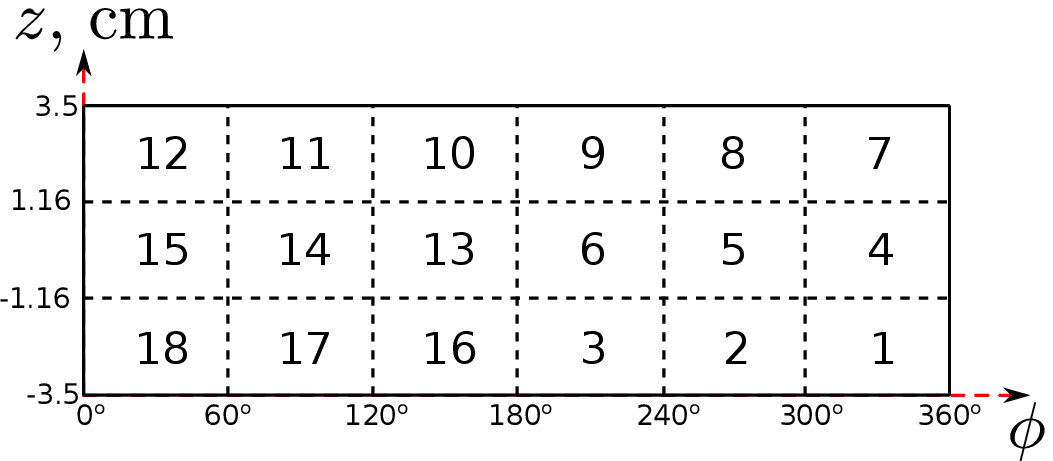}
  \end{minipage}
  \flushleft
  \hskip 3cm a)
  \hskip 6.5cm b)
\caption{(a) Coordinate system with the $x$ axis chosen to 
be on the 4--15 segment boundary.
Segment boundaries are indicated as dashed lines, coordinate axes as
dotted lines.
Numbers in parenthesis are segment numbers as
in (b). (b) Detector segment layout in ($\phi$,$z$) coordinates.}
\label{fig:det} 
\end{figure}

The detector used for this study was an 
n-type true-coaxial 18--fold segmented
detector with a diameter of 75\,mm, a height of 70\,mm and
an inner bore of 10\,mm. It was produced by Canberra, France.
The segmentation  scheme is three in $z$ and six in $\phi$.
The layout is depicted in Fig.~\ref{fig:det}.
The first such detector was characterized in detail \cite{char}
including the determination of the orientation of the crystallographic axes.
The second detector of the series, used for this study,  was 
previously used to test the performance of such a device while
submerged in a cryogenic liquid \cite{cold} and the temperature
dependence of pulse lengths \cite{T}.

The density of electrically active impurities, $\rho_{imp}$,  was
given  by the manufacturer as
$0.35\cdot10^{10}\mbox{cm}^{-3}$ at the top 
and $0.55\cdot 10^{10}\mbox{cm}^{-3}$ at the bottom of the detector.
The change in impurities is assumed to be linear with height, $z$.
The operational voltage was 2000\,V.
At the time of the measurements, the detector drew a leakage current
of up to 100\,nA, all through segment 9. 
It was, however, fully functional.
The only effect was that the energy resolutions in segment 9 and the 
core had deteriorated.

For all measurements,
the detector was mounted inside the vacuum cryostat K1 \cite{T}.
The detector was scanned in $\phi$ at $z=0$
with a $^{152}$Eu source placed at $r$=9.6\,cm.
It was also irradiated
from the top ($z=18.5$\,cm, r=0\,cm) with a $^{228}$Th and
a  $^{60}$Co source.
The $^{228}$Th source was also placed at the side 
($z=-2.3$\,cm, $r=17.6$\,cm, $\phi=155^\circ$), 
in front of segment 16.

\section{Detector scan}
\label{section:scan}

A well established way to find the crystallographic 
axes is to perform
an azimuthal scan, where
a source with a low energy gamma line, like $^{152}$Eu with its
122\,keV line, is moved around the 
detector in small steps in $\phi$. 
Low energy gamma rays predominantly deposit  their energy
close to the surface.
The rise times of the resulting pulses vary with $\phi$, reaching
a maximum for the drift along the $\langle 110 \rangle$ axes.

\begin{figure}[htbp]
\centering
  \includegraphics[width=0.95\textwidth]{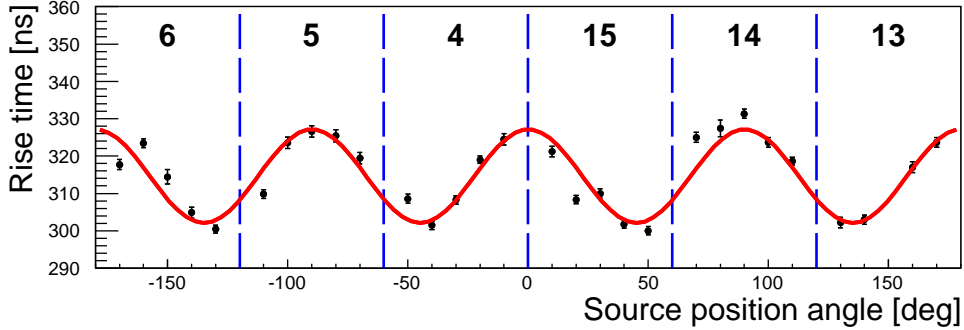}
\caption{
Dependence of rise time on the azimuth angle $\phi$ of the source position.
The points are the data with statistical uncertainties. If the 
uncertainty bars are not visible, they are smaller than the points. 
The line represents 
the fit described in the text. The vertical lines indicate the segment 
boundaries; the numbers are segment numbers.
}
\label{fig:scan} 
\end{figure}

A scan was performed with a collimated 
40\,kBq  $^{152}$Eu source with a 
1$\sigma$ beam spot diameter of about 5\,mm and a
step size of 10$^{\circ}$ in $\phi$.
The events were selected using an energy window of 10\,keV.
The rise time of a given pulse was determined by fitting a simulated pulse
to the measured one. This is described in detail 
elsewhere \cite{pss,T}. This method uses all the available 
information about a measured pulse while
suppressing the influence of noise. 
For every $\phi$ position, the average rise time,
$t_r$, was computed
for all pulses, for which
$\chi^2/{\rm{ndof}}<1.5$ for the fit with the simulated pulse.
Figure~\ref{fig:scan} shows the dependence of $t_r$
on the angle of the source position, $\phi$ . 
The data were fitted with  
the function 

\begin{equation}
t_r(\phi) = A+B\cdot \rm{sin}\left( \frac{2\pi}{90}\left(\phi+\phi_{\langle 110 \rangle}\right)\right) ~~,
\end{equation}

where A, B and $\phi_{\langle 110 \rangle}$ are the free parameters.
The coordinate system was chosen 
such that the 4--15 segment boundary was at $0^{\circ}$.

A study of the systematic uncertainties was performed. The energy window, in
which events were selected, was changed to 5\,keV and 20\,keV. In
addition, the $\chi^2$ used for the  pulse selection 
was modified from
1.5 to 1.2 and 1.8.
In both cases, 
a systematic uncertainty of 0.5$^{\circ}$ was deduced.
The dominating systematic uncertainty is connected to the
placement of the detector within the cryostat.
This could only be controlled to $3^{\circ}$.
The total systematic uncertainty was determined by adding the
individual contributions in quadrature.
The statistical uncertainty, as determined by the fit, 
is 0.4$^{\circ}$.
The final result is:
\begin{equation}
\phi_{\langle 110 \rangle} = -0.2^{\circ} \pm 0.4^{\circ} \rm{(stat.)}
                                       \pm 3.1^{\circ} \rm{(syst)} ~~.
\end{equation}

\section{Occupancy method}

The drift of the charge carriers within a germanium detector 
only follows radial lines along the crystallographic axes.
The transverse anisotropy causes bent trajectories elsewhere
in the crystal \cite[Fig.~3]{pss}.
As a result, the effective segment boundaries do not coincide with
the geometrical boundaries and the segment electrodes collect
charge carriers from volumina of different sizes.
This reflects in the occupancies of the segments when the detector 
is irradiated.

\subsection{Measured Occupancies}

\begin{figure}[htbp]
\flushleft
\hskip 3.6cm (a) \hskip 6.3cm (b)\\
\centering
  \includegraphics[width=0.48\textwidth]{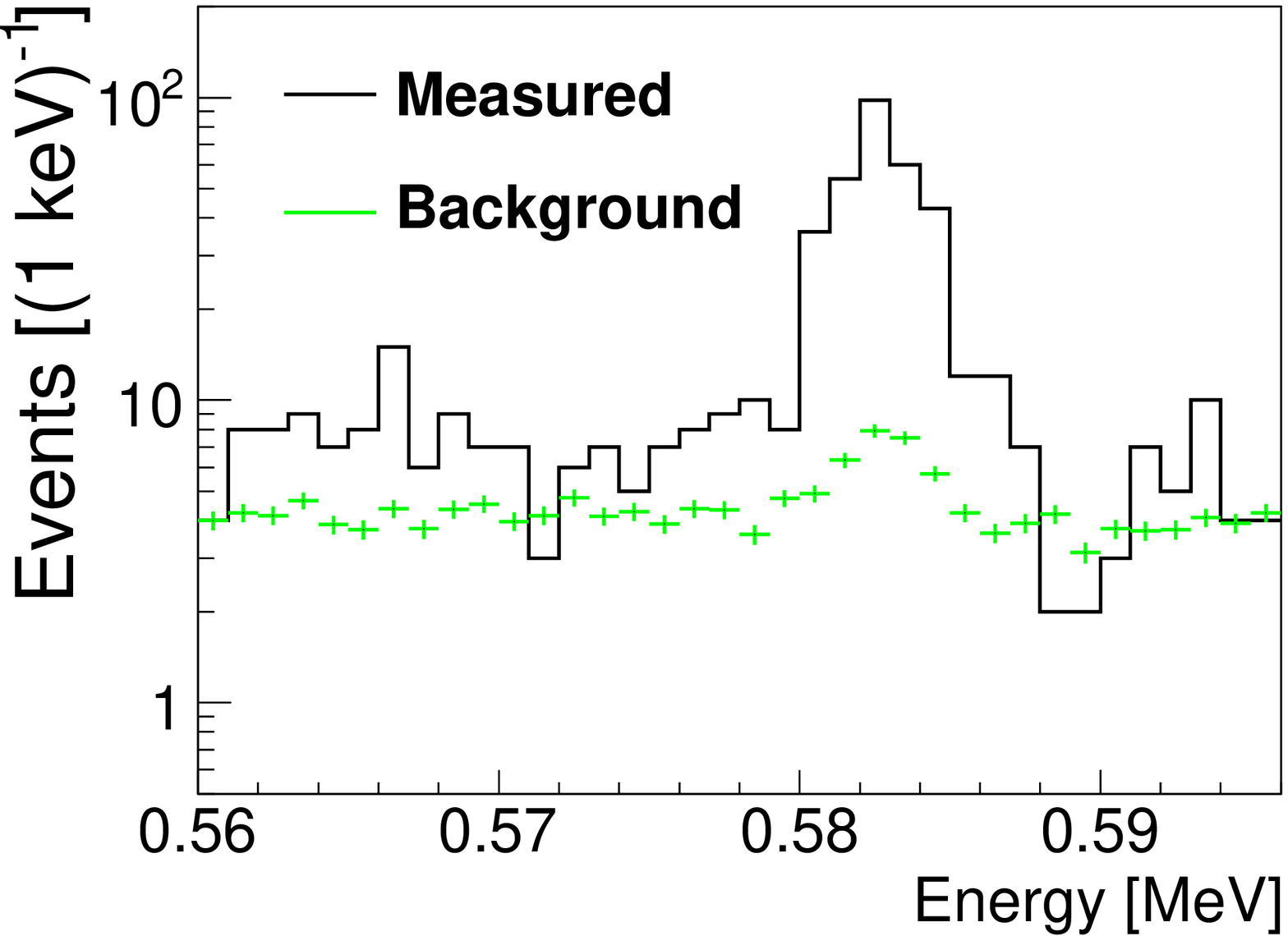}
  \includegraphics[width=0.48\textwidth]{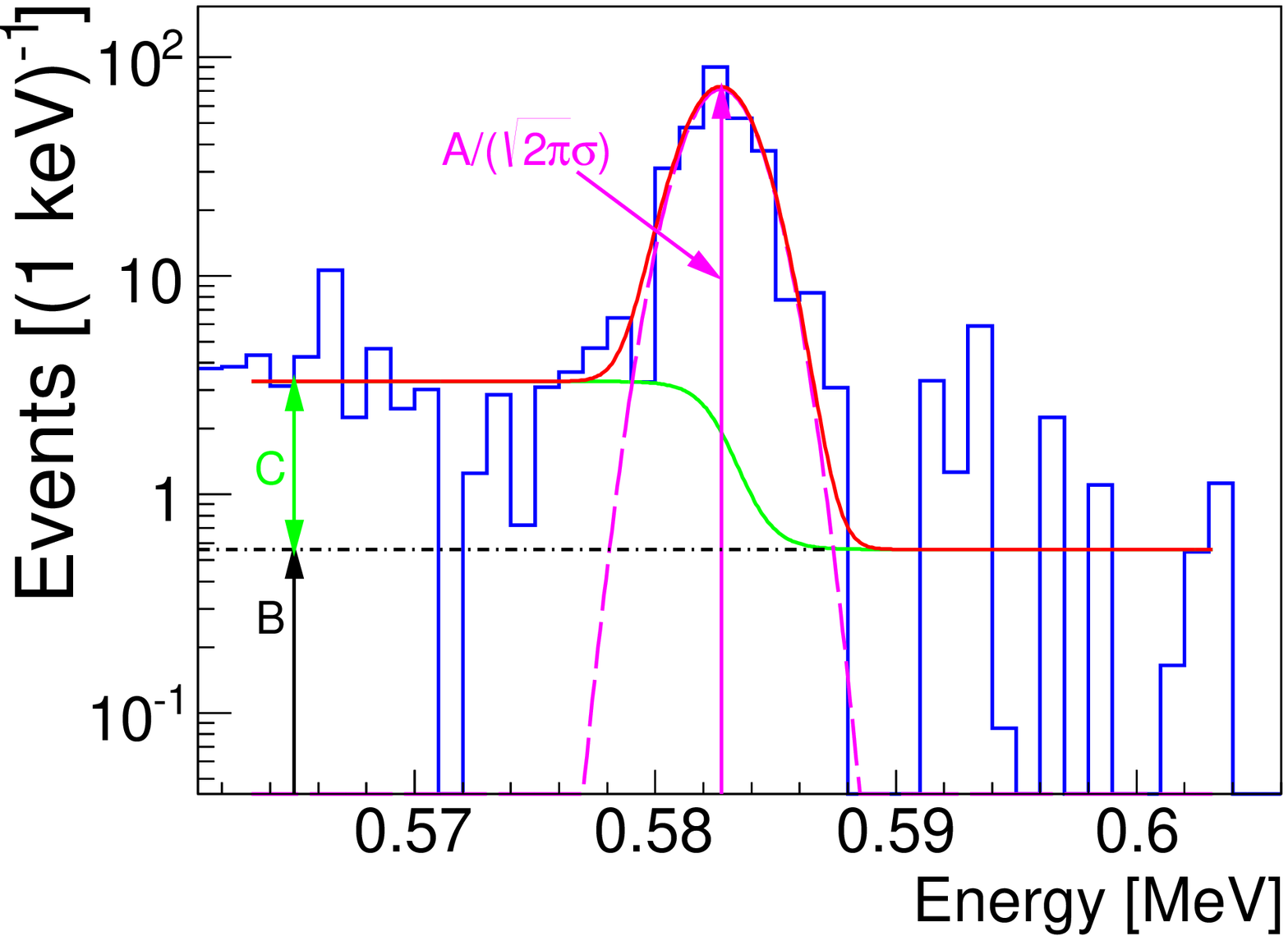}
\caption{
(a) Measured signal and background spectrum
in segment~3 around 0.58\,MeV and
(b) the corresponding fit to the background-subtracted measured spectrum
with the function given in Equation~\ref{eq:fits}.}
\label{fig:occ:fitcomponents}
\end{figure}

A 28\,kBq $^{228}$Th and a 40\,kBq $^{60}$Co source were used.
The occupancies were measured 
for the 0.58\,MeV and 2.61\,MeV lines from
$^{208}$Tl and the 1.17\,MeV and 1.33\,MeV lines from $^{60}$Co.
Only events, in which one segment
registered the same energy, $E_s$, as the core within 50\,keV, and 
the energy seen in any other segment was less than 150\,keV were used.
As the separately measured background spectra did not show lines
associated to $^{60}$Co, the background was not subtracted
for the analysis of these lines.
The background did show lines associated to $^{208}$Tl decays.
Therefore the background was subtracted for the analysis of the
corresponding spectra.
Figure~\ref{fig:occ:fitcomponents}(a) shows the spectrum 
measured in segment~3 for the irradiation from top and the
corresponding normalized background spectrum. The background contribution
for this relatively low-energy line is maximal at the bottom of the detector.
The resulting 
segment-spectra were fitted using the 
log-likelihood method. The fit function was a
combination of a Gauss function of width $\sigma$ and a sigmoid:

\begin{equation}
\label{eq:fits}
f(E_s)=\frac{A}{\sqrt{2\pi}\sigma} e^{-\frac{(E_s-E_\gamma)^2}{2\sigma^2}}+B+\frac{C}{e^{\frac{2(E_s-E_\gamma)}{\sigma}}+1}~~~,
\end{equation}

where $E_\gamma$ is the energy of the line under study and $A$, $B$, $C$
are the free parameters. 
The sigmoid represents the background shape. It drops for energies 
$E_s > E_\gamma$. The occupancy, $D_i$, of each segment $i$
was taken as the fitted
number of events under the peak,
$D_i=A\cdot{\rm{binwidth}^{-1}}$.

%--

\begin{figure}[htbp]
\centering
  \includegraphics[width=0.95\textwidth]{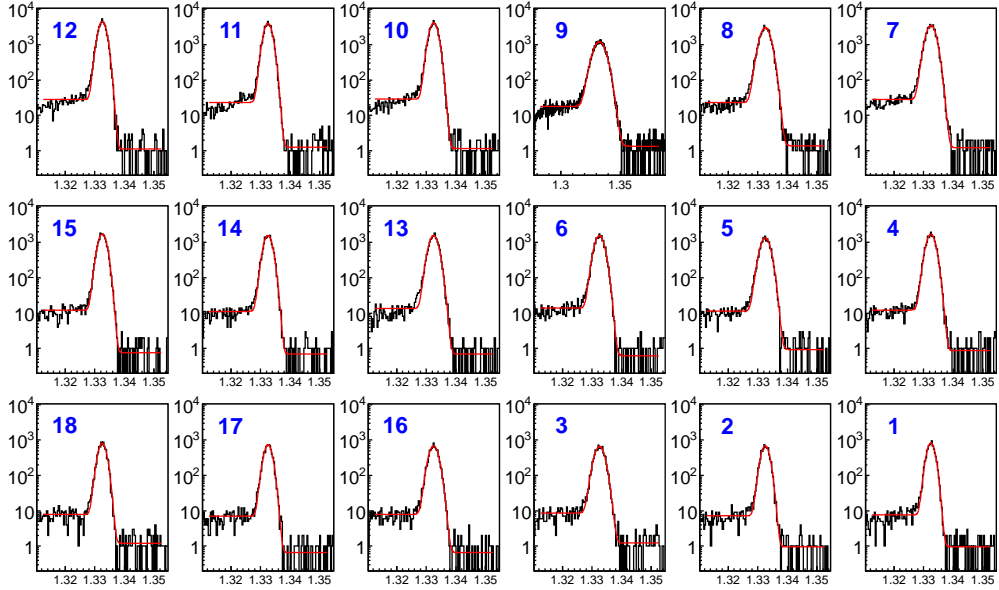}
\caption{
Energy spectra at $1.33$\,MeV as seen by the 18 segments
for the irradiation with $^{60}$Co from the top. Also shown are the
results of the fits with the function given in Equation~\ref{eq:fits}.}
\label{fig:occ:spectra}
\end{figure}

%--

\begin{figure}[htbp] 
\flushleft
\hskip 3.6cm (a) \hskip 6.3cm (b)\\
\centering 
  \includegraphics[width=0.48\textwidth]{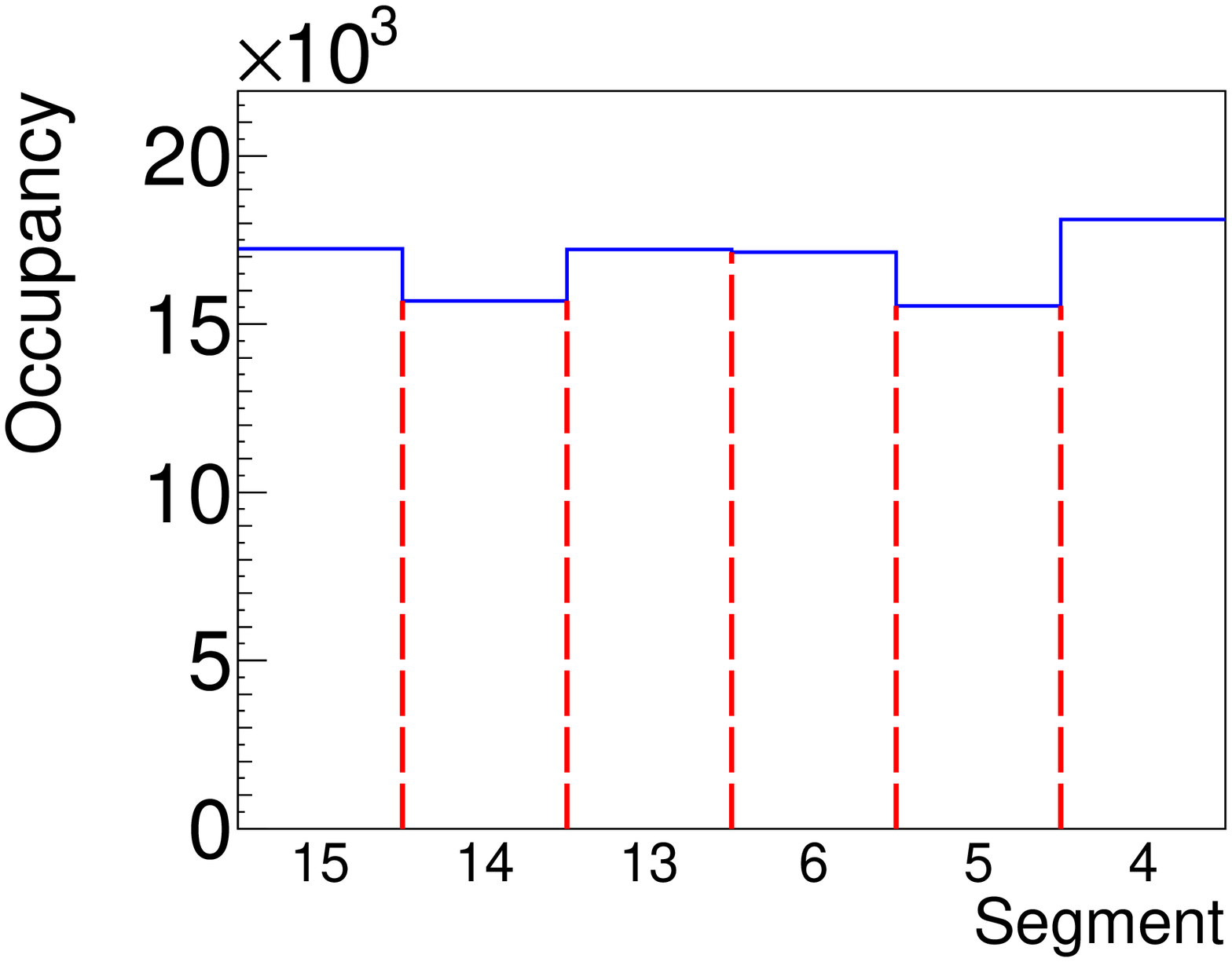}
  \includegraphics[width=0.48\textwidth]{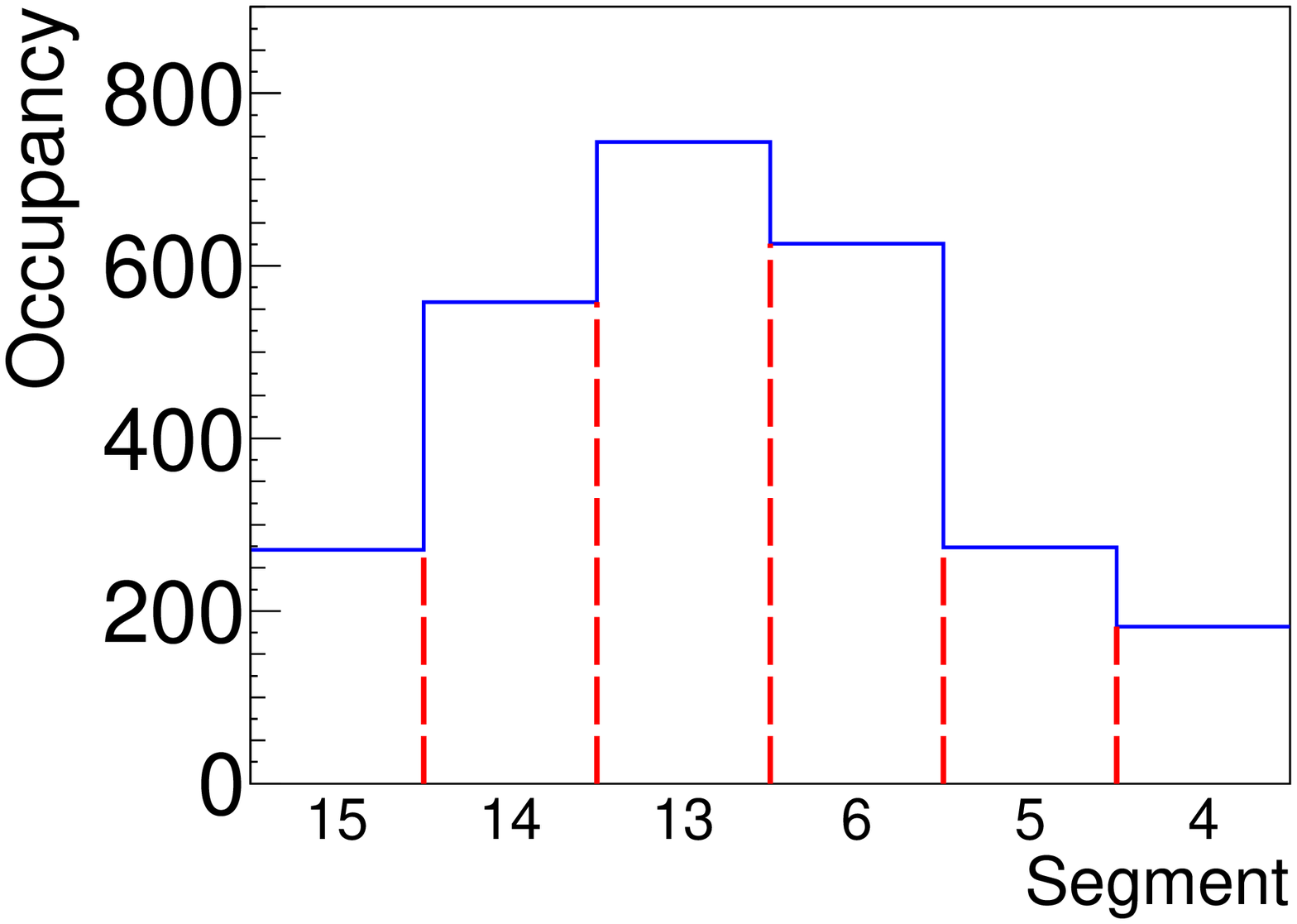}
\caption{
Measured occupancies (a) for the $1.33$\,MeV line extracted for the middle layer
for the irradiation from the top and (b) for the 2.61\,MeV line extracted 
for the middle layer for the irradiation from the side.
The numbers denote the segment numbers.}
\label{fig:occ:m133:s261} 
\end{figure} 

Figure~\ref{fig:occ:fitcomponents}(b) shows the result of the fit
to the background-subtracted spectrum around 0.58\,MeV
for the irradiation with $^{208}$Tl from
the top. The contributions from the terms associated to the parameters
$A,B$ and $C$ are indicated. The area in the peak associated 
to the parameter $A$ is well defined and
what enters the analysis.
Figure~\ref{fig:occ:spectra} shows the spectra of all 18 segments for
the irradiation with $^{60}$Co from the top \footnote{
Segment 9 in the top layer
has a significantly larger resolution and its occupancy is only used
in the analysis of data obtained by irradiation from the top.}.
Also shown are the fit results.

Figure~\ref{fig:occ:m133:s261}(a) 
shows the resulting occupancies for the middle
layer. A clear structure is visible. It is almost 
mirrored with respect
to the boundary between segments 13 and 6. This agrees with the 
naive expectation, because
the  $\langle 110 \rangle$ axis is 
almost aligned with this boundary.

Figure~\ref{fig:occ:m133:s261}(b) shows the measured occupancies 
for the 2.61\,MeV line in the middle layer
for the irradiation from the side. The situation is more complicated
in this case and there is no naive expectation
for the influence of the transverse anisotropy.
The occupancies 
were extracted for all layers and all energies and for both irradiation
from the top and the side.

\subsection{Expected Occupancies}
\label{section:expectations}

Events for the experimental setup
as described in Section~\ref{section:test_env}
were simulated with the \textsc{MaGe} \cite{MaGe} package and trajectories
were computed with the pulse shape simulation package \cite{pss}.
The package provides calculations of the electrical and weighting fields
and uses experimental input and model based calculations to provide
the mobilities, and thus the velocities, of holes and electrons at any given
point in the detector.
The orientation of the $ \langle 110 \rangle$ axis,
$\phi_{\langle 110 \rangle}^{\rm{sim}}$, was an input parameter to the
simulation of the drift.
The endpoints of the resulting
trajectories were used to assign each energy deposit of an event 
to a segment.
The energies were summed for each segment.
The event selection was performed as for the measurements.

\begin{figure}[htbp] 
\flushleft
\hskip 3.6cm (a) \hskip 6.3cm (b)\\
\centering 
  \includegraphics[width=0.95\textwidth]{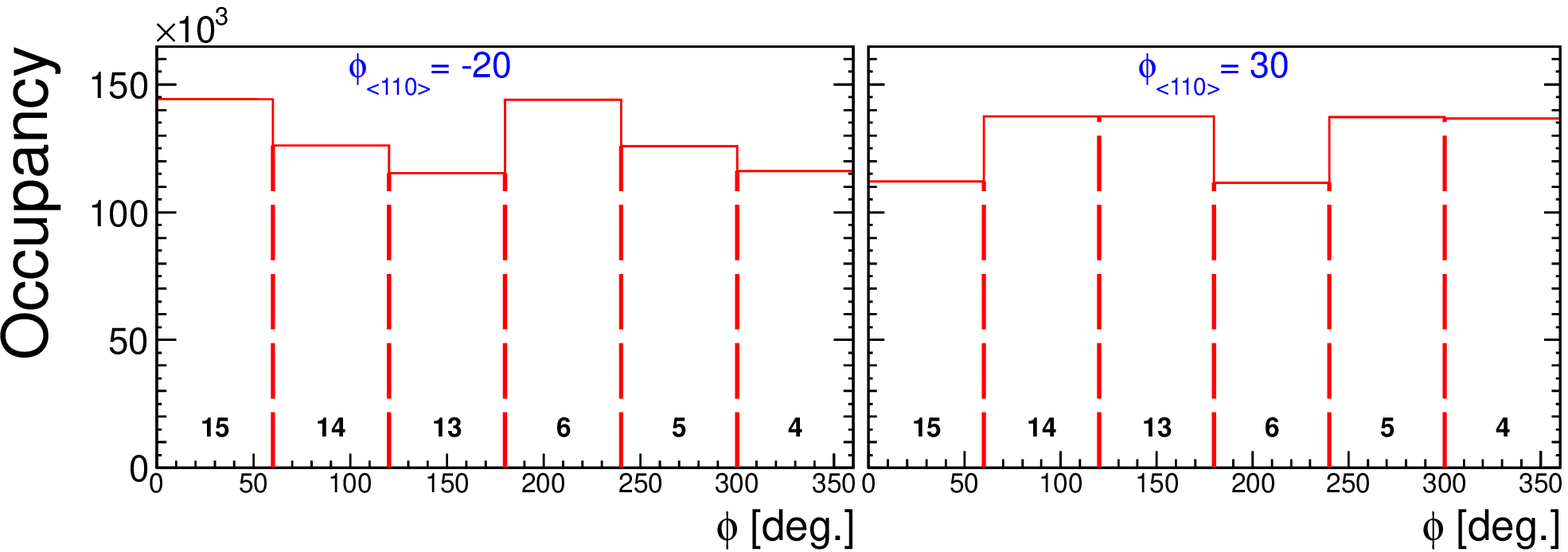}\\
  \includegraphics[width=0.95\textwidth]{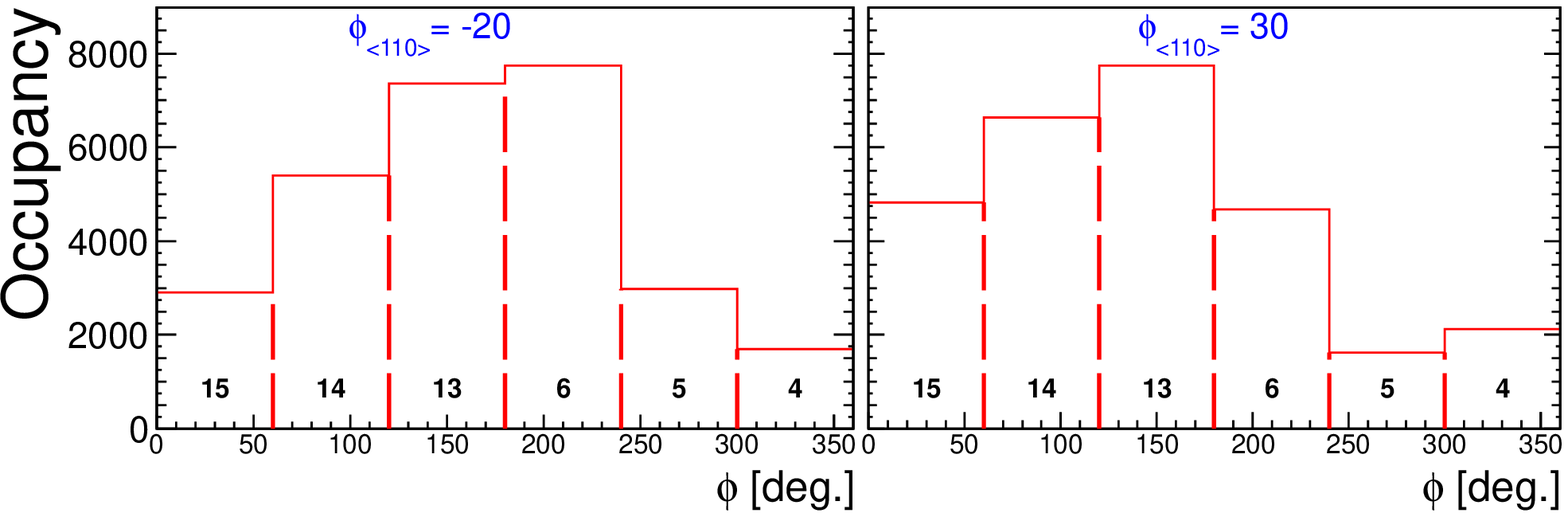}\\
\flushleft
\hskip 3.6cm (c) \hskip 6.3cm (d)\\
\caption{
Expected occupancy patterns 
in the middle layer for the irradiation
from the top/side for (a)/(c) 
$\phi_{\langle 110 \rangle}^{\rm{sim}}=-20^{\circ}$
and for (b)/(d) $\phi_{\langle 110 \rangle}^{\rm{sim}}=30^{\circ}$.
}
\label{fig:occ:sim} 
\end{figure}

The resulting expected occupancies for 
$\phi_{\langle 110 \rangle}^{\rm{sim}}=-20^{\circ}$ and 
$\phi_{\langle 110 \rangle}^{\rm{sim}}=30^{\circ}$
are shown in Fig.~\ref{fig:occ:sim} for the irradiation from the top and
the side. 
The occupancies depicted in Fig.~\ref{fig:occ:sim} 
were computed for the middle layer, for which an average
$\rho_{\rm{imp}}=0.45\cdot10^{10}\mbox{cm}^{-3}$ was assumed. 
For the irradiation from the top, the result for the 1.33\,MeV line 
of $^{60}$Co is shown. For the irradiation from the side, 
the 2.61\,MeV line of $^{208}$Tl was chosen.

For the irradiation from the top, the simulation 
of the middle layer predicts 
occupancy patterns which are almost 
 identical for segments 15--14--13
and 6--5--4. This reflects the 180$^\circ$ degeneracy of the 
axis orientation.
The mirror symmetry observed in data with respect to the 13--6 boundary 
disappears, if the  $\langle 110 \rangle$ axis is 
not aligned with this boundary.

For the irradiation from the side, a clear difference between the
predictions for $\phi_{\langle 110 \rangle}^{\rm{sim}}=-20^{\circ}$
and $\phi_{\langle 110 \rangle}^{\rm{sim}}=30^{\circ}$ is visible.
There are no apparent 
symmetries, but the relative occupancies significantly change
for varying 
$\phi_{\langle 110 \rangle}$.

\begin{figure}[htbp] 
\flushleft
\hskip 3.6cm (a) \hskip 6.3cm (b)\\
\centering 
  \includegraphics[width=0.95\textwidth]{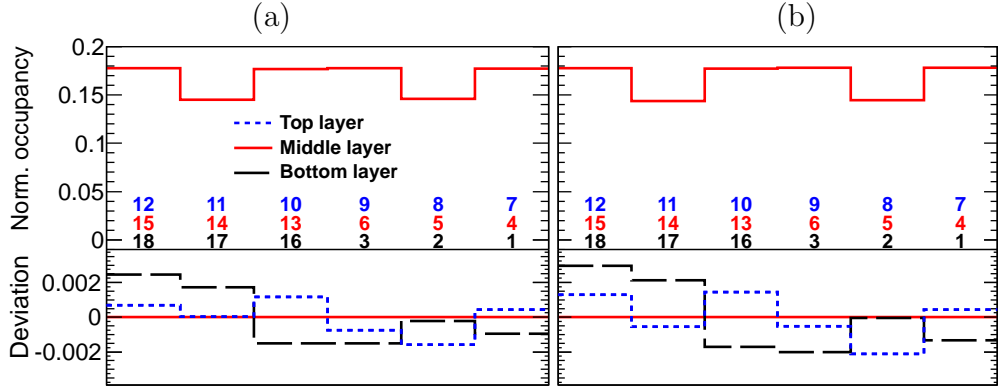}
\caption{
Comparison of the normalized
predicted occupancies for the $1.33$\,MeV line and 
$\phi_{\langle 110 \rangle}^{\rm{sim}}=0^{\circ}$ for 
the middle, top and bottom layers of (a) the nominal crystal 
and (b) a test crystal with lower $\rho_{imp}$. 
 }
\label{fig:sim:impdep} 
\end{figure} 

The amplitudes of the observed patterns depend on the amount
of transverse anisotropy in the crystal. This is influenced
by the hole mobility and by $\rho_{imp}$.
The dependence on $\rho_{imp}$ was investigated
by comparing expectations  for
the nominal detector, i.e. the one supposed to be close
to the real device with a $\rho_{imp}$ of 
%%%%%\begin{itemize} 
%\item nominal: 
$0.52\cdot10^{10} \mbox{cm}^{-3}$, 
$0.45\cdot10^{10} \mbox{cm}^{-3}$ and
$0.38\cdot10^{10} \mbox{cm}^{-3}$, 
%%%%%\end{itemize}
to predictions
for a lower impurity test detector with a $\rho_{imp}$ of 
%\begin{itemize} 
%\item test:
$0.29\cdot10^{10} \mbox{cm}^{-3}$, 
$0.18\cdot10^{10} \mbox{cm}^{-3}$ and
$0.06\cdot10^{10} \mbox{cm}^{-3}$ 
%%%%%\end{itemize}
for the top, middle and and bottom layers. 
Figures~\ref{fig:sim:impdep}(a) and (b) show the expectations
for  $\phi_{\langle 110 \rangle}^{\rm{sim}}=0^{\circ}$.

The amplitude of a pattern can be defined as the depth of the 
first step of the pattern, $A_{i-j}=S_i-S_j$, with $S_i$ and $S_j$ 
being the predicted occupancies for the segments $i$ and $j$ 
located in the same layer.
For the middle layer, 
$A_{15-14}=0.0326$ is predicted
for the nominal detector and 
$A_{15-14}=0.0336$ for 
the test detector.
This is a very small change of 0.6\,\% in a 20\,\% effect with 
respect to an occupancy of $S_i=1/6=0.17$ for all $i$, which
would be seen if there was no transverse anisotropy.

The Figs.~\ref{fig:sim:impdep}(a) and (b) also show predictions
for the top and bottom layers for the two simulated devices.
Shown are deviations, $\Delta_{k-l}=S_k-S_l$  with k and l 
being the relevant segment numbers in different layers, 
but at the same $\phi$.
The values of $\Delta_{k-l}$ reach 0.003, which is
a 2\,\% effect, i.e. 10\,\% of the amplitude.
The 15--14--13 6--5--4 degeneracy
is broken for the end layers due to a drift component in $z$, which
arises from $\rho_{imp}$ changing with $z$.
The $\Delta_{k-l}$
patterns are basically identical for the nominal and the test
device.

\begin{figure}[htbp] 
\flushleft
\hskip 3.6cm (a) \hskip 6.3cm (b)\\
\centering 
  \includegraphics[width=0.48\textwidth]{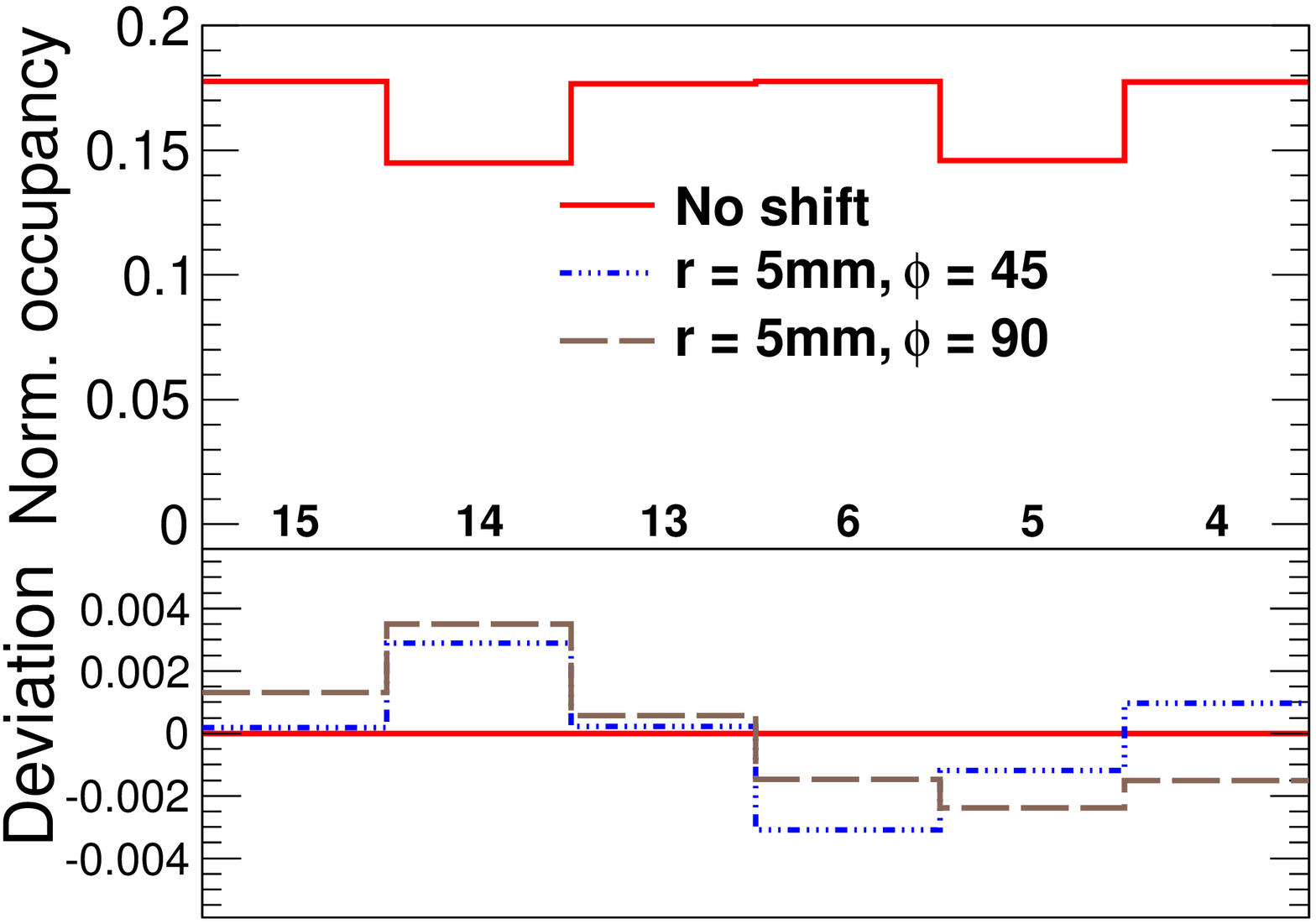}
  \includegraphics[width=0.48\textwidth]{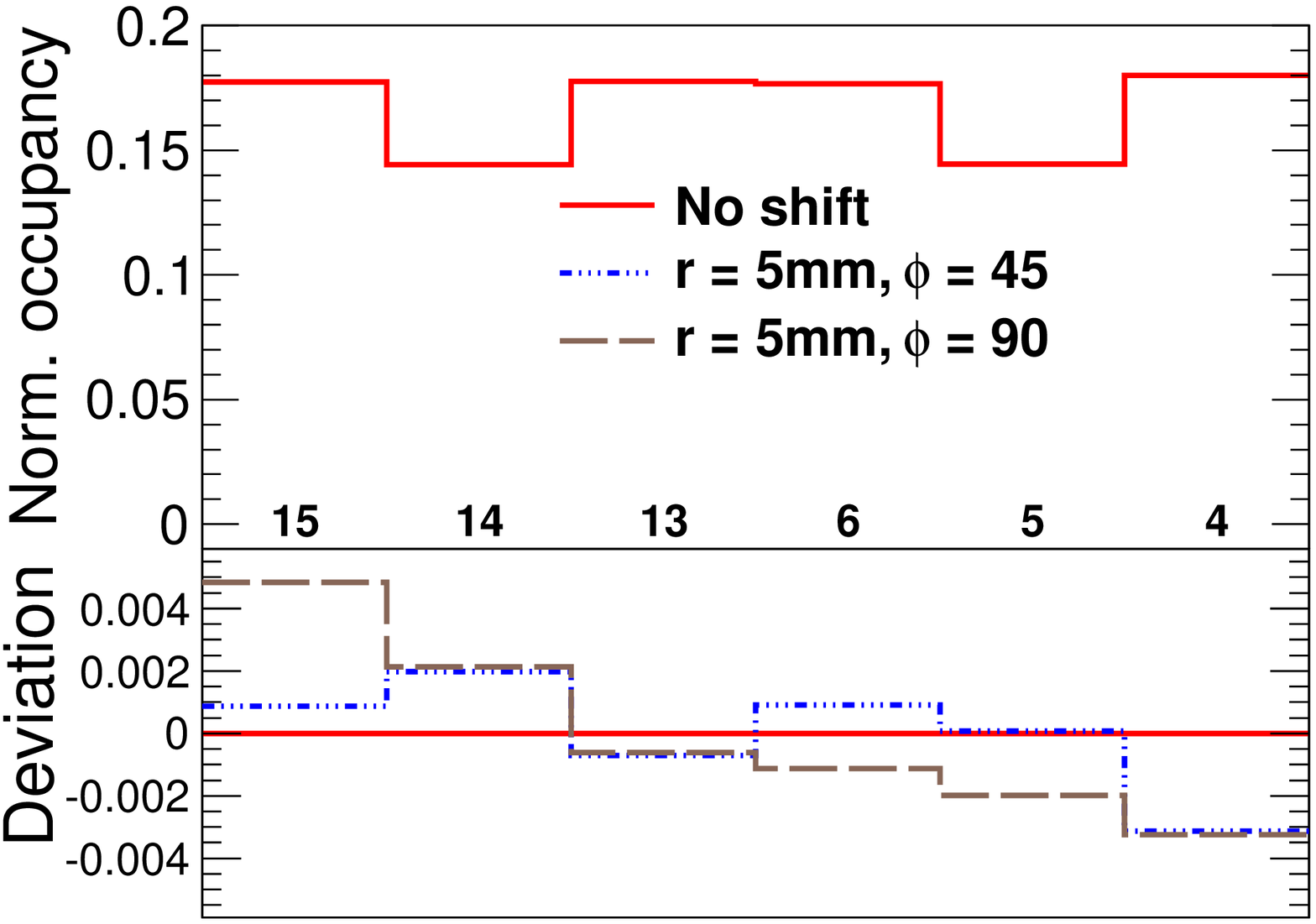}
\caption{
Expected occupancies for 
$\phi_{\langle 110 \rangle}^{\rm{sim}}=0^{\circ}$ 
for the middle layer  
for (a) the $1.33$\,MeV line 
and for
(b) the $2.61$\,MeV line for irradiation from the top.
Also shown are the changes 
predicted for a 5\,mm misalignment of the
source along 45$^\circ$ or 90$^\circ$. 
}
\label{fig:sim:imploc} 
\end{figure}

The sensitivity of the predictions 
on the exact location of the source is
demonstrated in Fig.~\ref{fig:sim:imploc} for the case of (a) the 1.33\,MeV
line of $^{60}$Co and (b) the 2.61\,MeV line of $^{208}$Tl.
Shown are the normalized predicted occupancies
for a perfectly aligned source and the effects on the pattern 
for two source misalignments of 5\,mm,
one along $\phi=45^\circ$ and one along  $\phi=90^\circ$.
The misalignments cause changes in the pattern 
for the middle layer breaking the 15--14--13 
6--5--4 degeneracy. 
The effect on a single occupancy is as much as 0.005, which
is a 3\,\% effect on the occupancy, i.e. as much as 
15\,\% on the amplitude.

\subsection{Extraction of the axes orientation}

The occupancies were computed for 
$\phi_{\langle 110 \rangle}^{\rm{sim}}$ 
varying in 1$^\circ$ steps.
The resulting occupancies were analyzed by computing
the test statistic $\epsilon$, defined as

\begin{equation}
  \label{eq:axes_determ:goodness}
  \epsilon=\sum_{i} \frac{(D_i-S_i)^2} {D_i^2} ~~~,
\end{equation}

where $D_i$ and $S_i$ denote the measured and simulated 
occupancies in segment $i$. The simulation was normalized to the
total number of events in a given layer and
the sum runs over all segments in this layer.

\begin{figure}[htbp] 
\flushleft
\hskip 3.6cm (a) \hskip 6.3cm (b)\\
\centering 
  \includegraphics[width=0.48\textwidth]{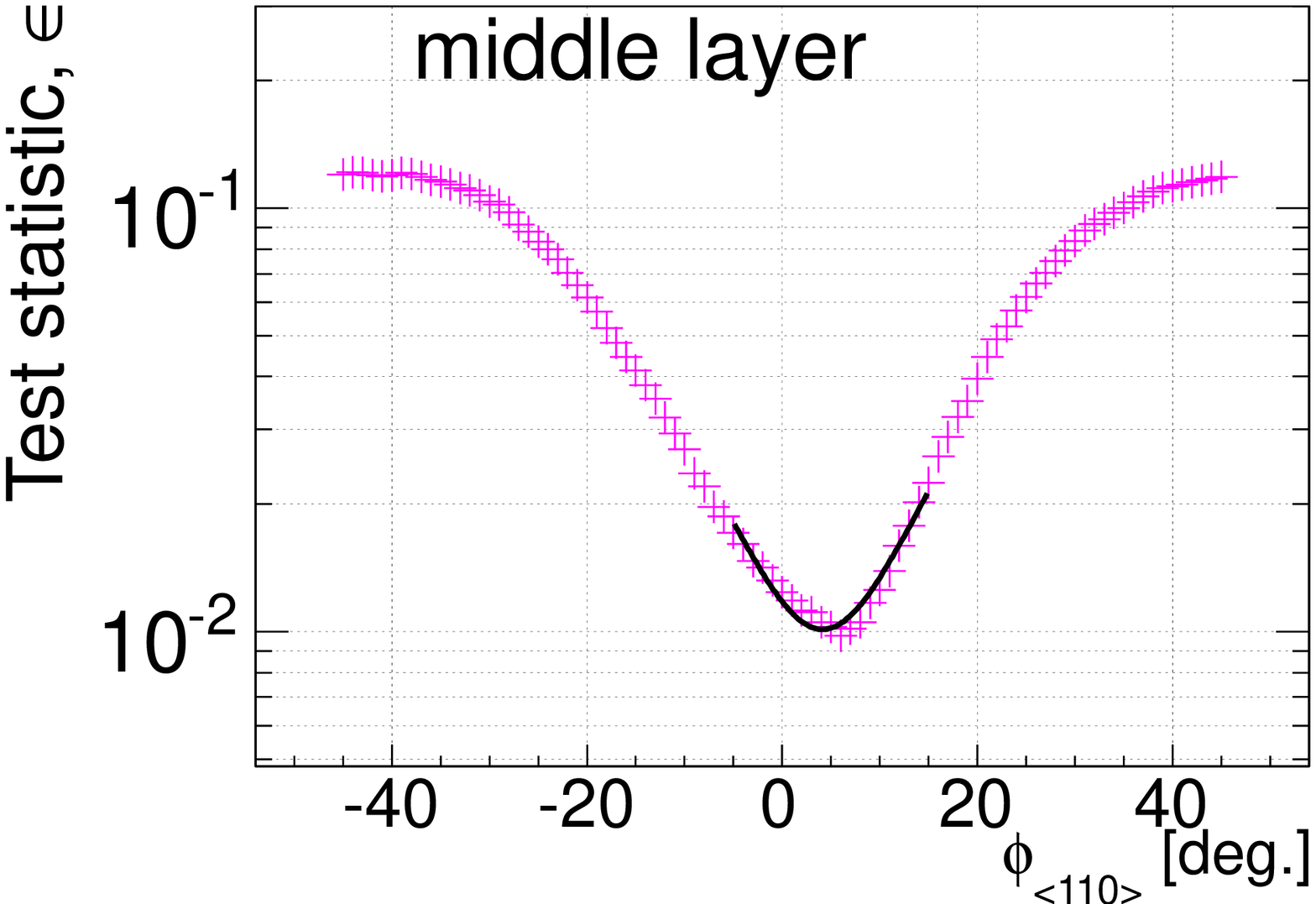}
  \includegraphics[width=0.48\textwidth]{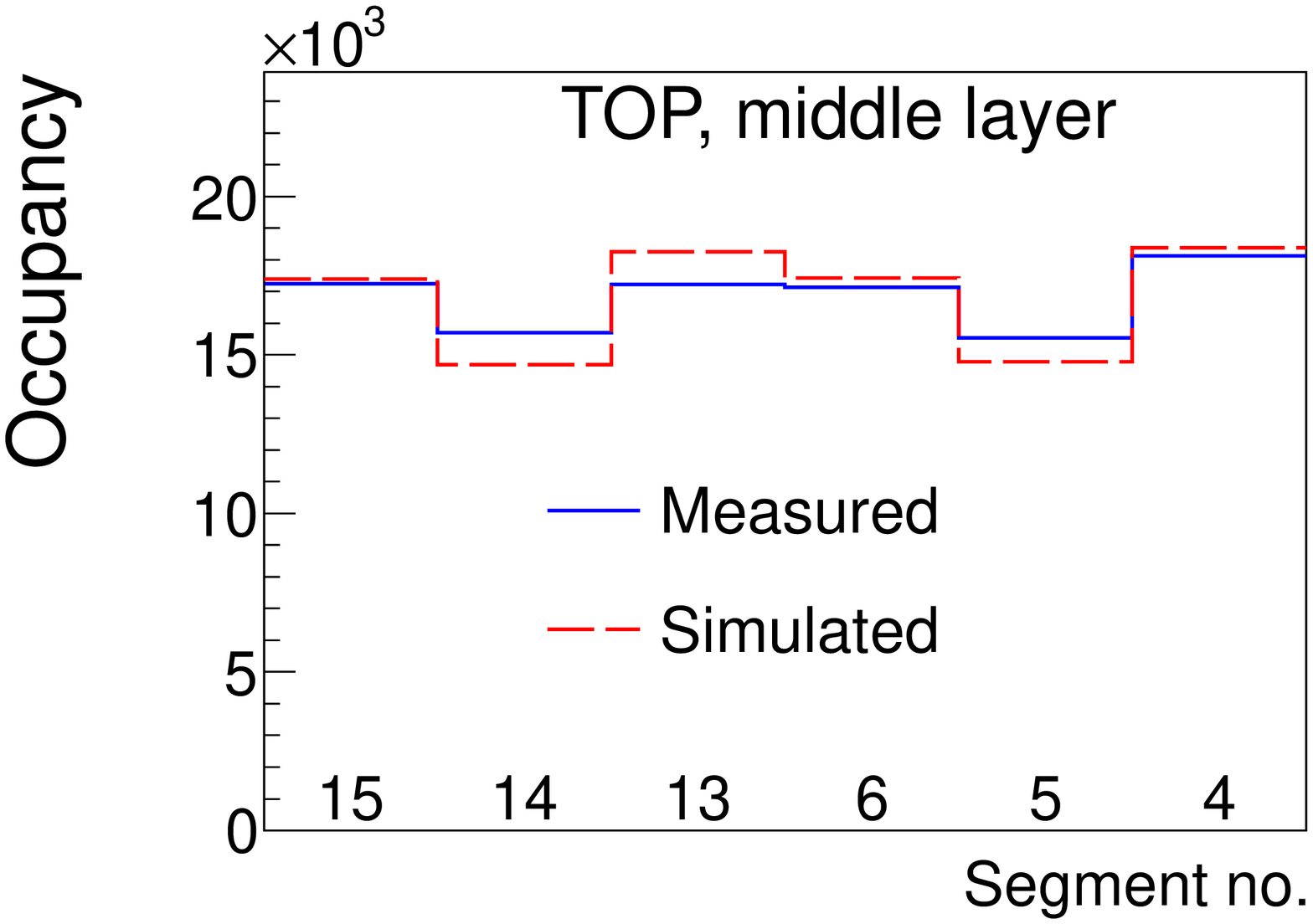}
\caption{
(a) The dependence of $\epsilon$ on $\phi_{\langle 110 \rangle}^{\rm{sim}}$
for the $1.33$\,MeV line in the middle layer
for the irradiation from the top. (b) The corresponding
measured and expected occupancies.
The numbers denote the segment numbers. The simulation was normalized
to the observed number of events in the middle layer.}
\label{fig:eps:ft113} 
\end{figure}

The resulting function $\epsilon(\phi_{\langle 110 \rangle}^{\rm{sim}})$ 
was fitted in a 20$^\circ$ window 
with a second order polynomial.
The $\phi_{\langle 110 \rangle}^{\rm{sim}}$ corresponding
to the minimum
of the polynomial, $\epsilon_{\rm{min}}$, 
was taken as the result of the procedure, $\phi^{\rm{meas}}$.
Figure~\ref{fig:eps:ft113}(a)
shows  $\epsilon(\phi_{\langle 110 \rangle}^{\rm{sim}})$
plus the result of the polynomial fit
for the case of the
1.33\,MeV line in the middle layer for irradiation from the top. 
The shape of the pattern observed in the data is best reproduced by the
prediction for 
$\phi_{\langle 110 \rangle}^{\rm{sim}} \approx4^\circ$.
Figure~\ref{fig:eps:ft113}(b) depicts
the corresponding measured and predicted occupancies.

The comparison between the predicted and measured
pattern in Fig.~\ref{fig:eps:ft113} shows that 
the shape of the pattern is very well reproduced. 
However, the amplitude of the
predicted pattern is larger than
the amplitude of the measured pattern.
As was shown in Section~\ref{section:expectations},
this cannot be explained 
by the choice of 
$\rho_{imp}$ in the simulation or
by any misalignment of the source. 
As the hole mobility is not well known \cite{bart},
its simulation could be the source of the
discrepancy.

\begin{figure}[htbp] 
\flushleft
\hskip 3.6cm (a) \hskip 6.3cm (b)\\
\centering 
  \includegraphics[width=0.48\textwidth]{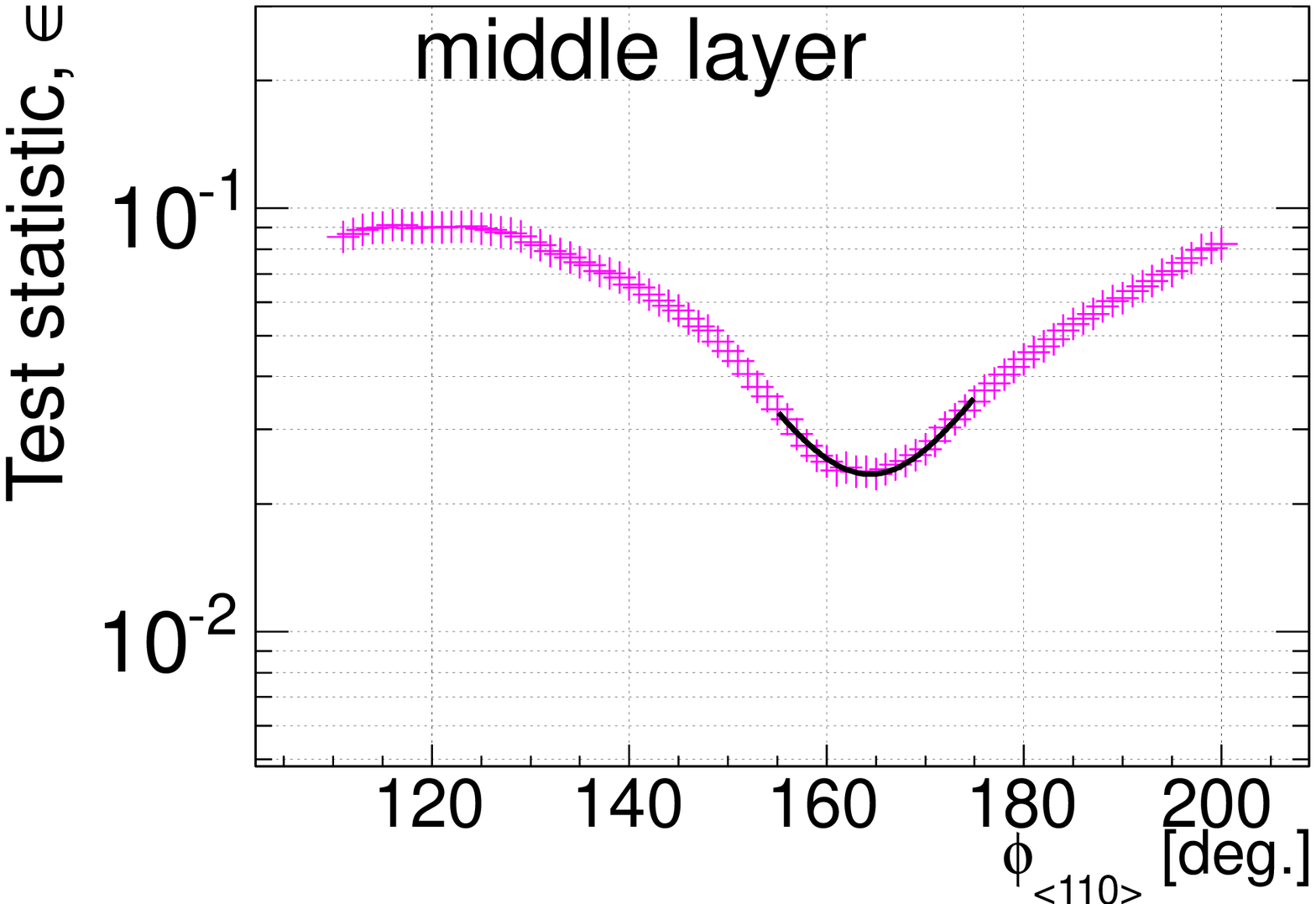}
  \includegraphics[width=0.48\textwidth]{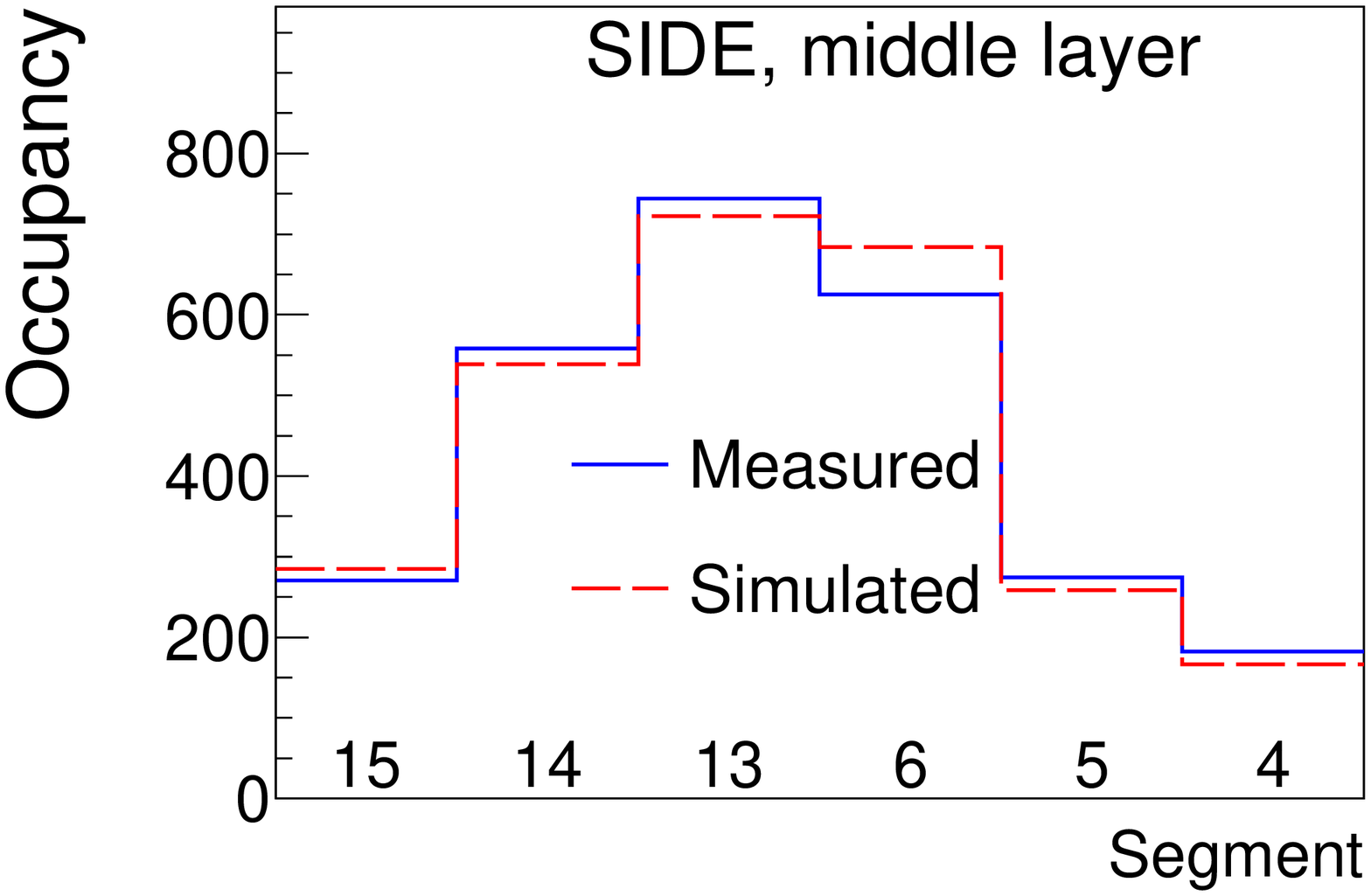}
\caption{
(a) The dependence of $\epsilon$ on $\phi_{\langle 110 \rangle}^{\rm{sim}}$
for the $2.61$\,MeV line in the middle layer
for the irradiation from the side. (b) The corresponding
measured and expected occupancies.
The numbers denote the segment numbers. The simulation was normalized
to the observed number of events in the middle layer.}
\label{fig:eps:fs261} 
\end{figure}

Figure~\ref{fig:eps:fs261} shows the situation for the 2.61\,MeV line
in the middle layer for the irradiation from the side.
The value of
$\epsilon_{\rm{min}}$ is significantly higher than for the top case
depicted in
Fig.~\ref{fig:eps:ft113}, but the minimum is as distinct.
The pattern observed in the data is well reproduced by the
prediction for $\phi_{\langle 110 \rangle}\approx164^\circ\widehat{=}-16^\circ$.

\subsection{Systematic Uncertainties}
The systematic uncertainties fall into two categories:
\begin{enumerate} 
\item parameters used in the treatment of the data
\item imperfections of the simulation
\end{enumerate}

The first category is dominated by uncertainties arising
through the definition of 
$\epsilon$ and the fit to the resulting $\epsilon$-function.
The test statistic $\epsilon$ deviates from a classical $\chi^2$
by putting extra emphasis on bins with lower occupancies.
For the irradiation from the top, the differences 
in the results are small
and the resulting systematic uncertainty is 1$^\circ$.
For the irradiation from the side, the differences 
in the results are significant.
This is intended. However, a study using different test statistics
yielded a systematic uncertainty of 4$^\circ$.
The effect of a variation of the width of the fit window 
for the $\epsilon$-function was also studied.
The window was widened to 30$^\circ$
and narrowed to 10$^\circ$ and the resulting uncertainty is 1$^\circ$.
Using other functions then a second order polynomial to fit
the $\epsilon$-function yielded a systematic uncertainty of 2$^\circ$.

The total systematic uncertainty connected with $\epsilon$ is
evaluated by adding the individual uncertainties in quadrature.
It is 2.5$^\circ$ and 4.6$^\circ$ for irradiation from the
top and the side, respectively.

In the second category, the influence of the combination of hits
before drifting the charge carriers was studied. Per default, hits within
a distance of 1\,mm were clustered. That was changed to 2\,mm and the
uncertainty due to this was found to be negligible.
Any diffusion of the charge cloud is not simulated; the effect is
expected to be small compared to the influences of charge clustering.

The main uncertainties are connected with
the assumptions on the detector parameters
and the source location.

As shown in Section~\ref{section:expectations},
a mismatch of simulated and real $\rho_{imp}$ slightly changes
the amplitude of the pattern, but not the shape.
Therefore, the absolute values of $\epsilon$ will change, but the
result will actually not. This was confirmed by simulation.

Also discussed in Section~\ref{section:expectations}
was the influence of any misalignment of the source. 
The source position was controlled to about 5\,mm. 
Different layers are 
affected slightly differently due to the difference in angular coverage.
The simulations show that the systematic uncertainty due to possible
misalignments of up to 5\,mm is 5$^\circ$ for irradiation from the top.

The systematic uncertainty for the irradiation from the side
directly reflects  the uncertainty in the $\phi$-position of the source
with respect to the detector. The detector was remounted for the
side  measurement and its relative rotation within the cryostat
could, as mentioned before in Section~\ref{section:scan}, only be controlled to
3$^\circ$. In addition, the position of the source relative to
the cryostat could only be controlled to 5$^\circ$.
Added in quadrature, this results in a total systematic uncertainty
of 6$^\circ$.

For the analysis of lines which are also present in the background,
possible changes in the background over time are another source of
systematic uncertainty.
Such changes can occur, if the cryostat or any other equipment
in the lab is moved. Unfortunately, this was the case.
The resulting patterns are very sensitive,
especially for the irradiation from the side.
The uncertainty was evaluated conservatively by performing the complete
analysis for the $^{208}$Tl lines with and without background subtraction.
The result is a systematic uncertainty
of 1$^\circ$ and 7$^\circ$ for the top and side cases, respectively.

In total, the systematic uncertainties for the irradiation
from the top and the side were evaluated to be 
6$^\circ$ and 10$^\circ$,
respectively. They are dominated by the uncertainty 
in the source position and,
for the irradiation from the side, possible changes in the background.

\subsection{Results}

The axes orientation was determined for all energies and all layers.
The results, $\phi^{\rm{meas}}$, 
are given in Table~\ref{tab:res:all}.
Also given are weights, $w=\epsilon^{-1}_{\rm{min}}$, for all measurements.
The statistical uncertainty of the 
fit to the $\epsilon$-function was always
negligible compared to the systematic uncertainty.

\begin{table}[!h]
   \centering
   \begin{tabular}{c|r|r|r|r|r|r}
     \hline
     \hline
     E$_{\rm{line}}$ & \multicolumn{2}{|c|}{Top layer} & 
\multicolumn{2}{|c|}{Middle layer} & \multicolumn{2}{|c}{Bottom layer} \\
     \cline{2-7}
  [MeV] & $\phi^{\rm{meas}}$ & \multicolumn{1}{c|}{\it w} & 
$\phi^{\rm{meas}}$ & \multicolumn{1}{c|}{\it w}& 
$\phi^{\rm{meas}}$& \multicolumn{1}{c}{\it w}\\
     \hline
     \hline
\multicolumn{7}{l} {Irradiation from the top:} \\ 
     \hline
 0.58 & -11.1$^\circ$ & 79  & -8.6$^\circ$ & 52 & 13.1$^\circ$ & 52 \\
 1.17 &  -7.7$^\circ$ & 100 &  1.4$^\circ$ & 115& -3.7$^\circ$ & 167 \\
 1.33 &  -7.1$^\circ$ & 88  &  4.2$^\circ$ & 99 & 1.1$^\circ$ & 117 \\
 2.61 &  0.3$^\circ$ & 38  & -6.2$^\circ$ & 23 & -6.3$^\circ$ & 44 \\
     \hline
 \multicolumn{7} {l} {Irradiation from the side:} \\ \hline
 0.58 & -21.9$^\circ$ & 33 & -14.2$^\circ$ & 52 & -23.4$^\circ$ & 63\\
 2.61 & -18.6$^\circ$ & 19 & -15.6$^\circ$ & 43 & -15.9$^\circ$ & 24 \\

     \hline
   \end{tabular}
   \vskip 0.2cm
   \caption{Extracted $\langle 110 \rangle$-axis positions, 
    $\phi^{\rm{meas}}$,
    for different gamma lines and layers. 
    Also listed are the weights, $w$, as defined in the text.}
   \label{tab:res:all}
\end{table}

For the irradiation from the top, the results for 
the $^{208}$Tl lines show a spread 
which is slightly larger than expected
from the evaluation of the systematic uncertainties. 
This could be due to the effect of statistical fluctuations
in individual bins.
An ensemble test was used to determine the statistical 
uncertainties.
The test was performed by 
randomly varying the content of each bin
according to its statistical uncertainty and extracting
$\phi^{\rm{meas}}$ for the modified occupancy pattern.
The resulting $\phi^{\rm{meas}}$ distribution
was fitted with a Gaussian function and the result used to
determine the statistical uncertainty.
The largest uncertainties were observed  
for the 2.61\,MeV line for the irradiation from the top.
They were 4$^\circ$
for all layers.

The statistical and systematic uncertainties
are different for different energies and 
layers. Therefore, weighted averages were computed:

\begin{equation}
  \label{eq:axes_determ:sum_weights}
  \phi_{\langle 110 \rangle}^{\rm{meas}}=
   \frac{\sum \phi_{\rm{min}}^{\rm{meas}}\cdot w} {\sum w}~~,
\end{equation}
where the sums run over a combination of layers and line energies.
The results for the $^{60}$Co and $^{208}$Tl lines in the three layers
for irradiation from the top are:

\begin{equation}
\phi_{\langle 110 \rangle}^{\rm{Co,top}} = 
       -1.8^{\circ} \pm 1^{\circ} {\rm{(stat)}}
       \pm 6^{\circ} {\rm{(syst)}} ~~,
\end{equation} 
\begin{equation}
\phi_{\langle 110 \rangle}^{\rm{Tl,top}} = 
       -3.6^{\circ} \pm 4^{\circ} {\rm{(stat)}}
       \pm 6^{\circ} {\rm{(syst)}} ~~.
\end{equation}

\begin{figure}[htbp] 
\flushleft
\hskip 3.6cm (a) \hskip 6.3cm (b)\\
\centering 
  \includegraphics[width=0.48\textwidth]{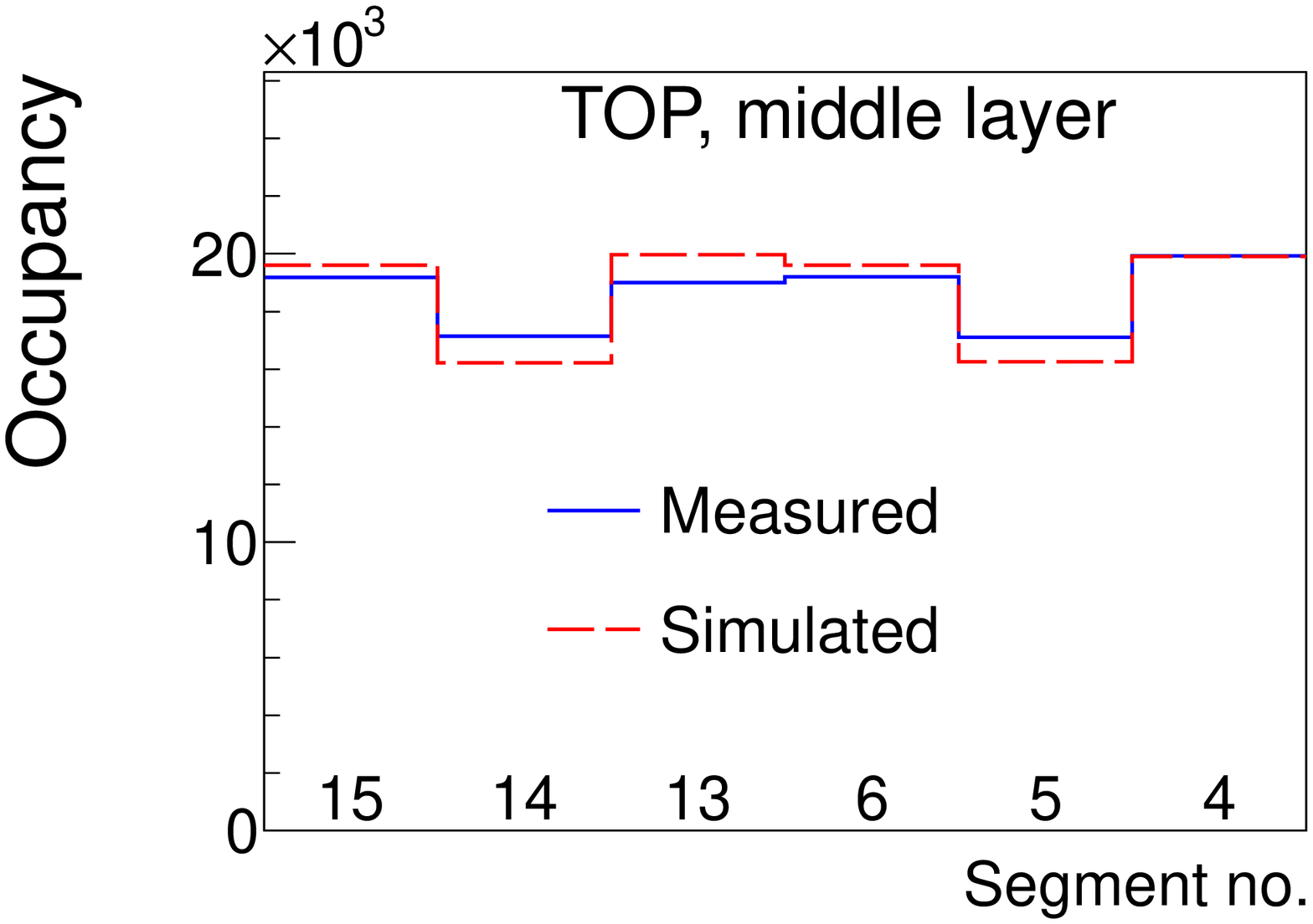}
  \includegraphics[width=0.48\textwidth]{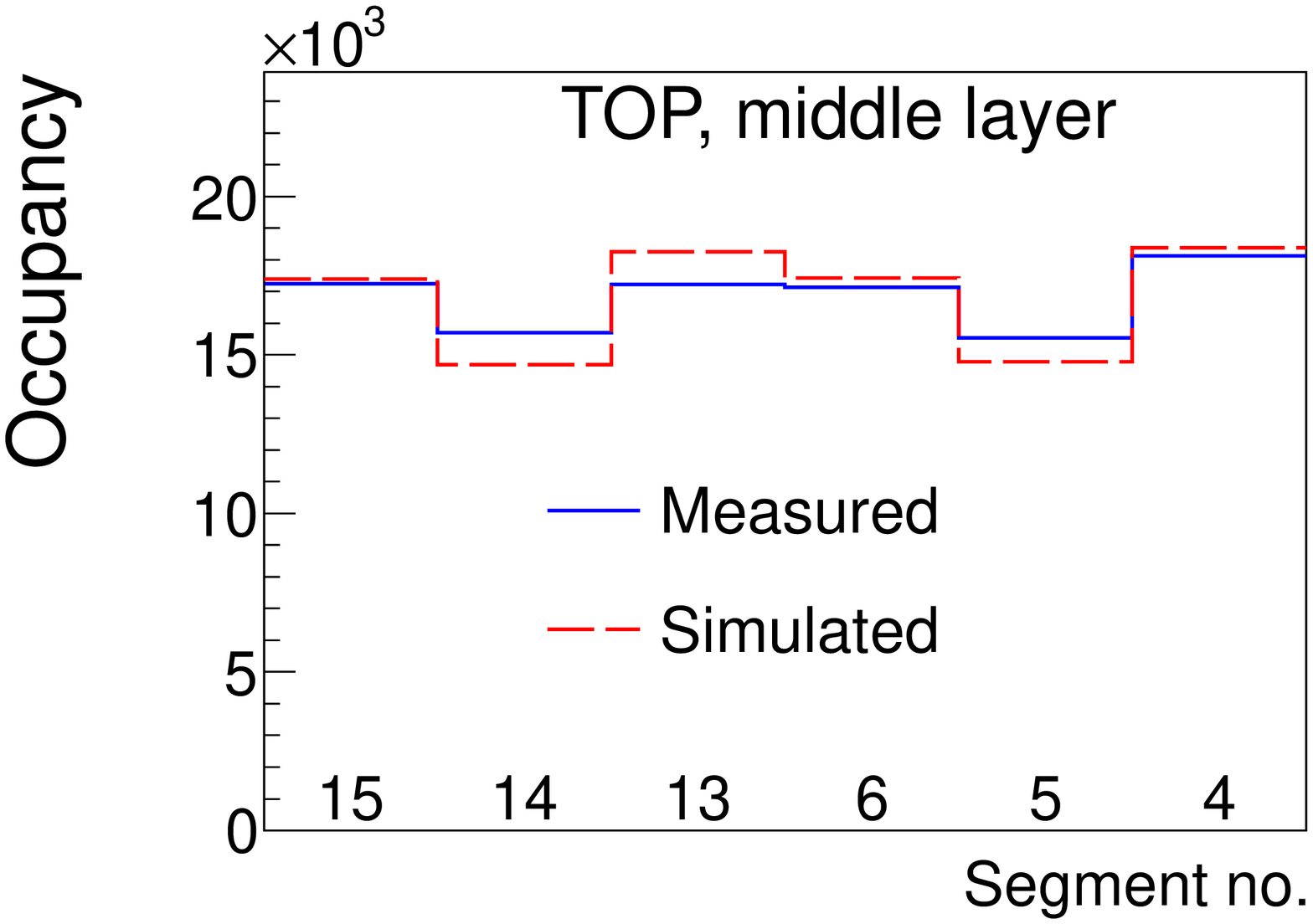}\\

  \includegraphics[width=0.48\textwidth]{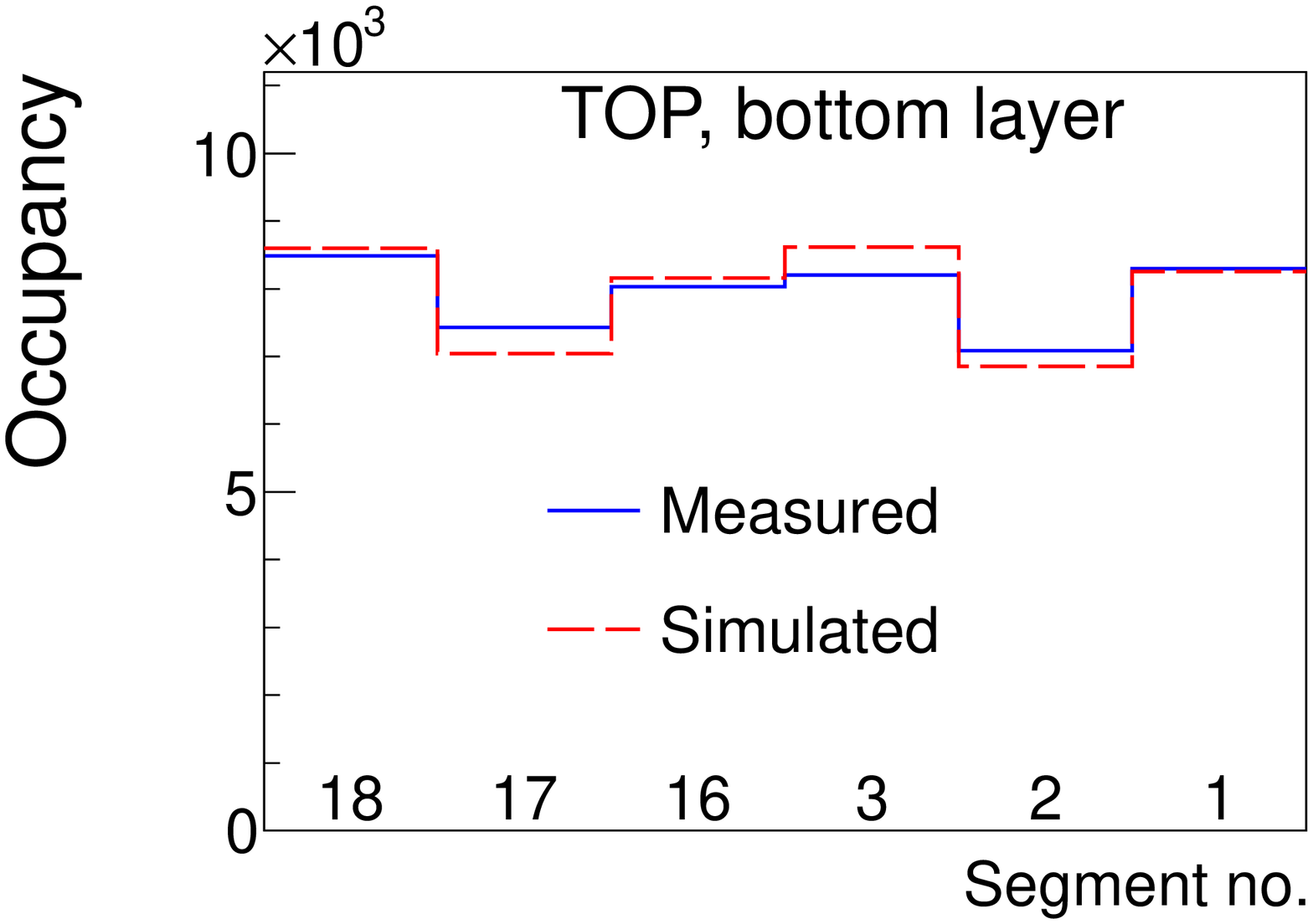}
  \includegraphics[width=0.48\textwidth]{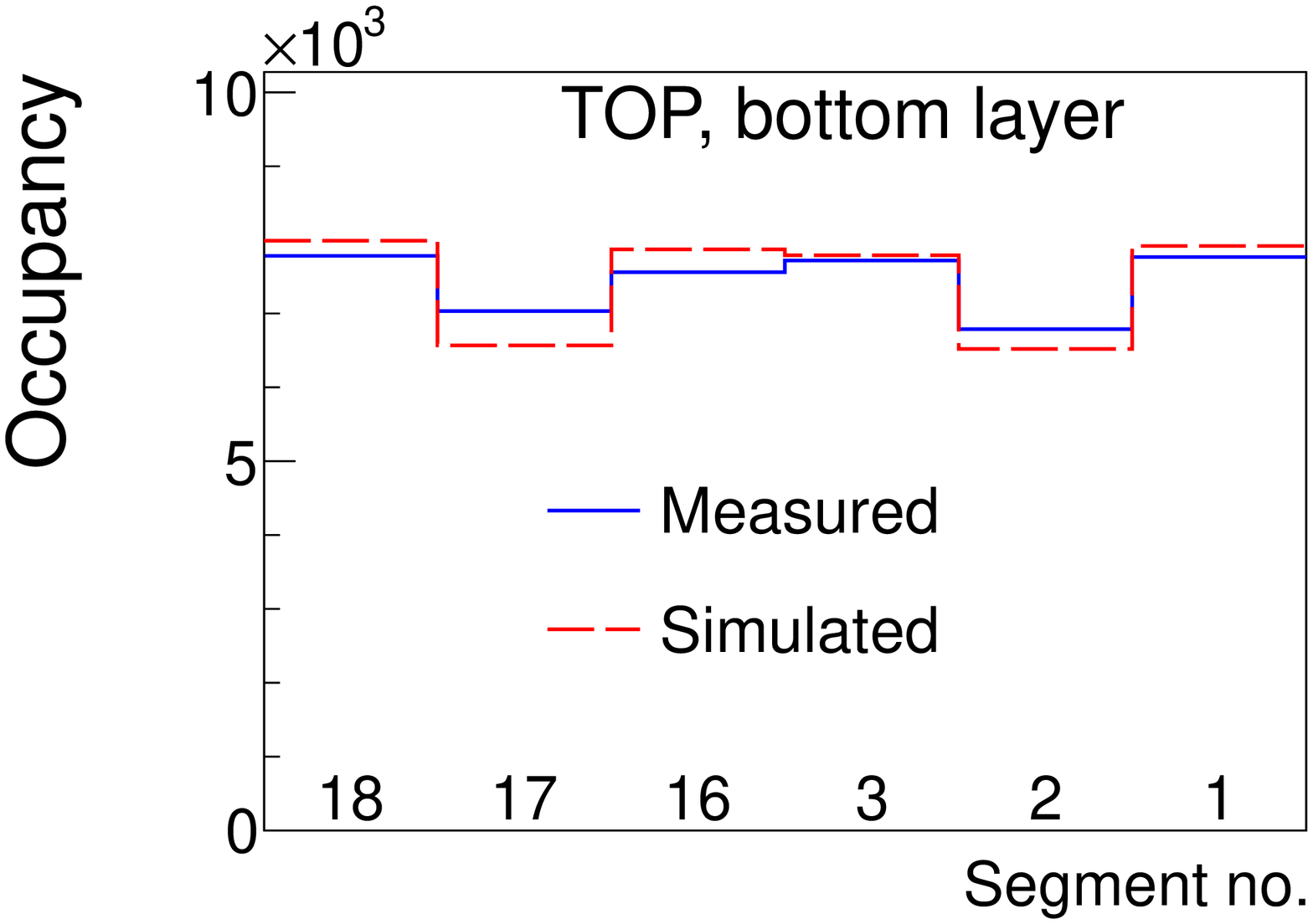}
\flushleft
\hskip 3.6cm (c) \hskip 6.3cm (d)\\
\caption{
Measured occupancies together with the best predictions for
the irradiation with $^{60}$Co from the top 
for (a) 1.17\,MeV, middle layer
(b) 1.33\,MeV, middle layer
(c) 1.17\,MeV, bottom layer
(d) 1.33\,MeV, bottom layer.
}
\label{fig:occ:res:tCo} 
\end{figure}

Figure~\ref{fig:occ:res:tCo} 
shows the measured occupancies together with the best predictions
for the middle and bottom layer for the $^{60}$Co irradiation.
The high statistics $^{60}$Co distributions show 
that the predictions reproduce the shapes of all four
measured patterns well, but that the predicted amplitudes 
are consistently too large.
The measured amplitudes are about 40\,\% smaller than the
predicted ones.
As discussed previously, 
this cannot be explained by an impurity level assumed to be too high
in the simulation. 
The reproduction of the shapes of the measured distribution 
also indicates that the alignment of the source was well within 
the 5\,mm uncertainty.

The simulation uses the hole mobilities as 
calculated by the pulse shape simulation package \cite{pss} 
using measured input parameters \cite{bart}.
There is a non-negligible uncertainty in these calculations and the
input parameters. The observation of less 
pronounced patterns than predicted could eventually be used to 
test the assumptions about hole mobilities.
A simulation confirmed that the discrepancy between the predicted
and observed amplitudes could be explained by an underestimate
of the ratio of the mobilities along $\langle 111\rangle$ and 
$\langle 100\rangle$.
The trajectories were recalculated for mobilities,
for which the component along the $\langle 111\rangle$ axis
was globally,
i.e. independent of the electric field, increased by 10\,\%.
This resulted in amplitudes about
the size observed.
However, as the calculation of mobilities along
the $\langle 110\rangle$ axis and between the axes depend on
several approximations, this will need further 
systematic studies.

\begin{figure}[htbp] 
\flushleft
\hskip 3.6cm (a) \hskip 6.3cm (b)\\
\centering 
  \includegraphics[width=0.48\textwidth]{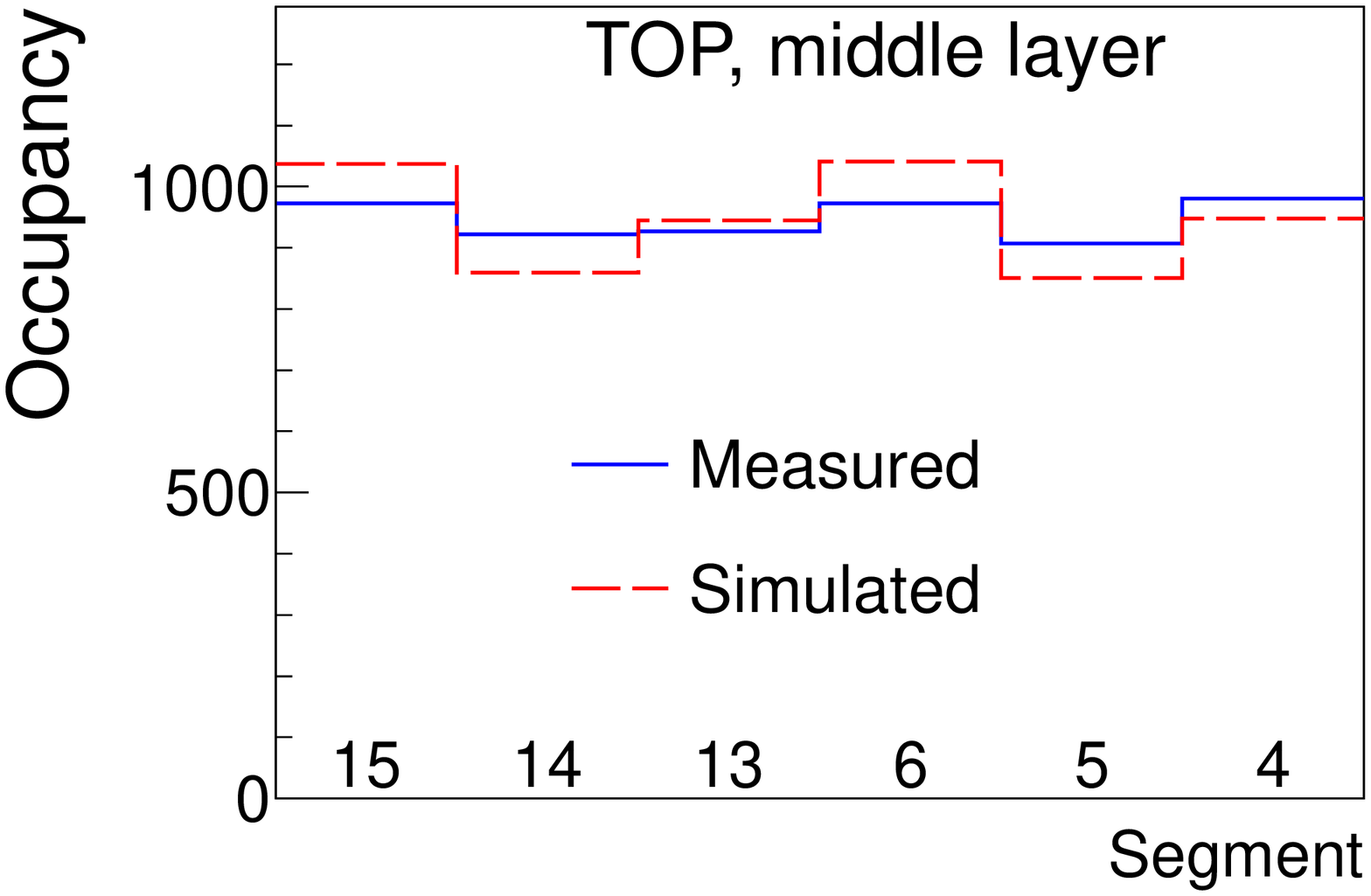}
  \includegraphics[width=0.48\textwidth]{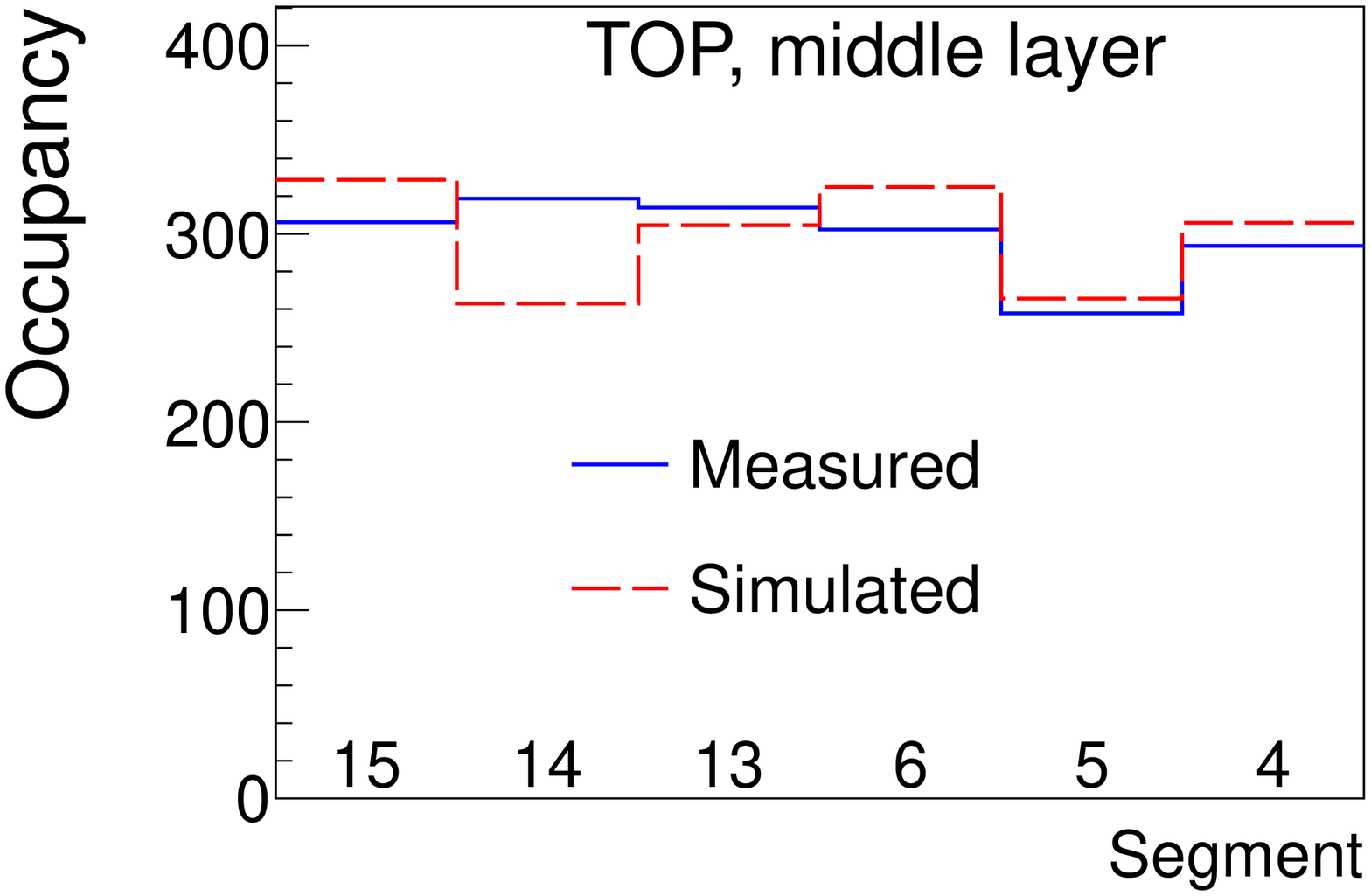}\\

  \includegraphics[width=0.48\textwidth]{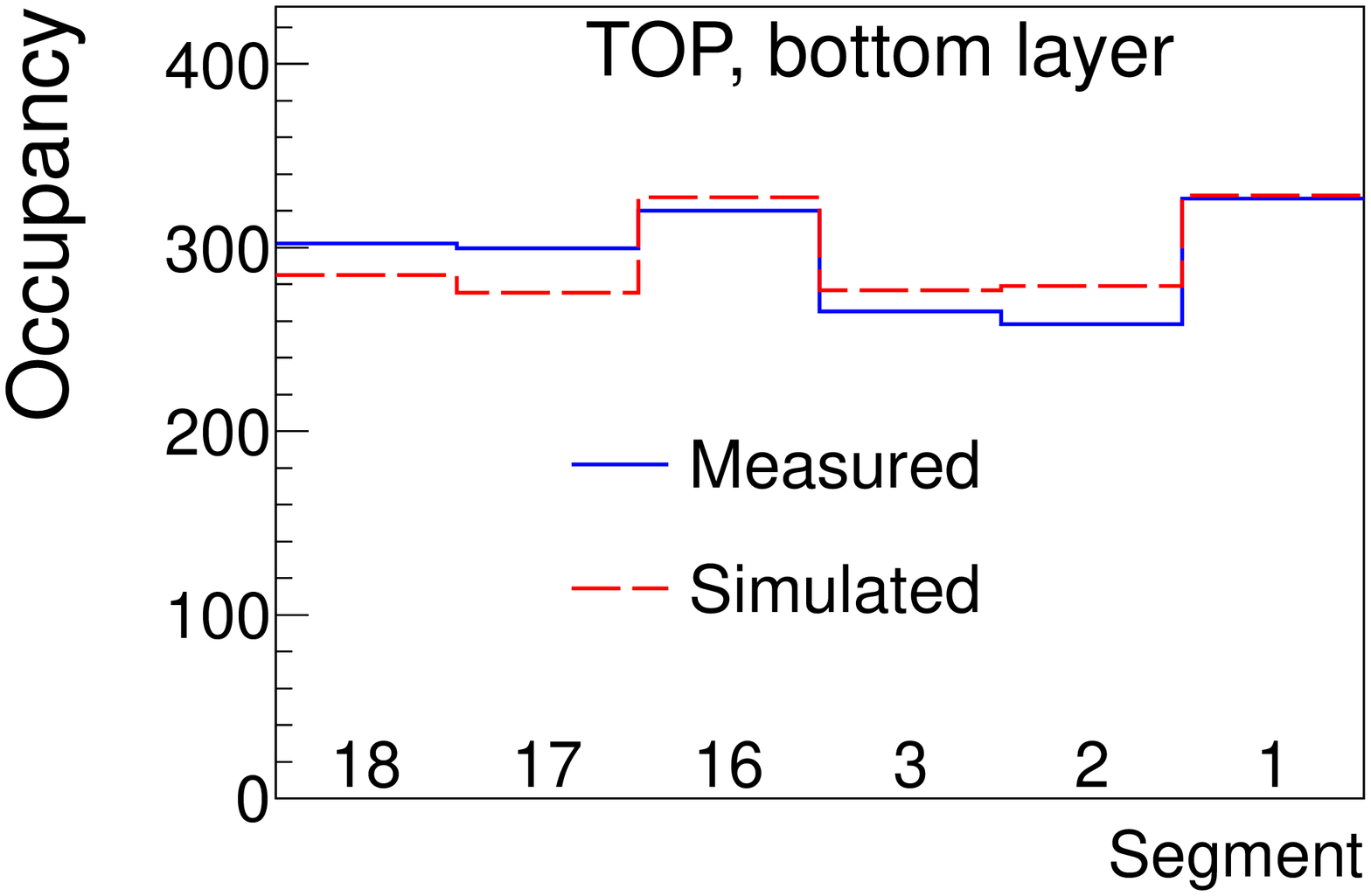}
  \includegraphics[width=0.48\textwidth]{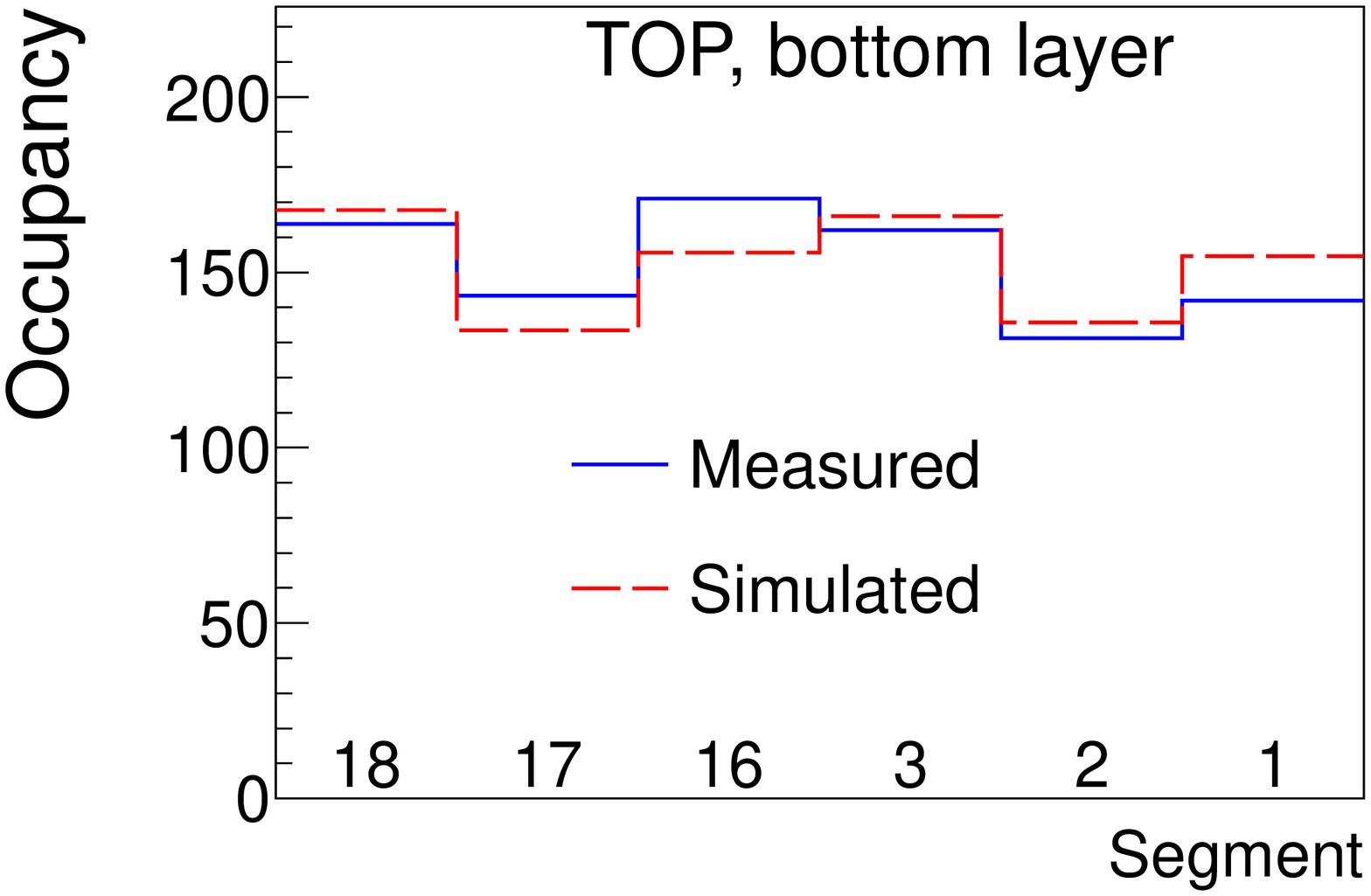}
\flushleft
\hskip 3.6cm (c) \hskip 6.3cm (d)\\
\caption{
Measured occupancies together with the best predictions for
the irradiation with $^{208}$Tl from the top 
for (a) 0.58\,MeV, middle layer
(b) 2.61\,MeV, middle layer
(c) 0.58\,MeV, bottom layer
(d) 2.61\,MeV, bottom layer.
}
\label{fig:occ:res:tTh} 
\end{figure}

Figure~\ref{fig:occ:res:tTh} shows the measured occupancies 
together with the best predictions
for the middle and bottom layer for the $^{208}$Tl irradiation
from the top.
The statistics is much lower than for 
the irradiation with $^{60}$Co, such that single bins can be affected
by significant statistical fluctuations.
The 2.61\,MeV line does not have a large 
probability to get fully absorbed
in the volume of a single segment. 
The bottom layer has fewer events by design, because of the
absorption in the crystal.
It should be noted that the spread of the single measurements 
is still
reasonable, even though single bins are fluctuating.

%-----

%-----

The individual values obtained for the irradiation from the side
deviate consistently by more than 10$^\circ$ from the expectation.
The ensemble test for the 2.6\,MeV line,
as discussed previously in this
chapter, 
yielded the statistical uncertainty of 4$^\circ$ 
and 6$^\circ$ for the middle and bottom layers, respectively.
Segment~9 was not used in the measurements
using the top layer. Even though this segment was directly facing
the source, the resulting measurements are still consistent with
the results for the other layers.
The top layer was, however, excluded from the averaging.
The overall result is:

\begin{equation}
\phi_{\langle 110 \rangle}^{\rm{Tl,side}} = 
       -17.9^{\circ} \pm 5^{\circ} {\rm{(stat)}}
       \pm 10^{\circ} {\rm{(syst)}} ~~.
\end{equation}

\begin{figure}[htbp] 
\centering 
\hskip 3.6cm (a) \hskip 6.3cm (b)\\
  \includegraphics[width=0.48\textwidth]{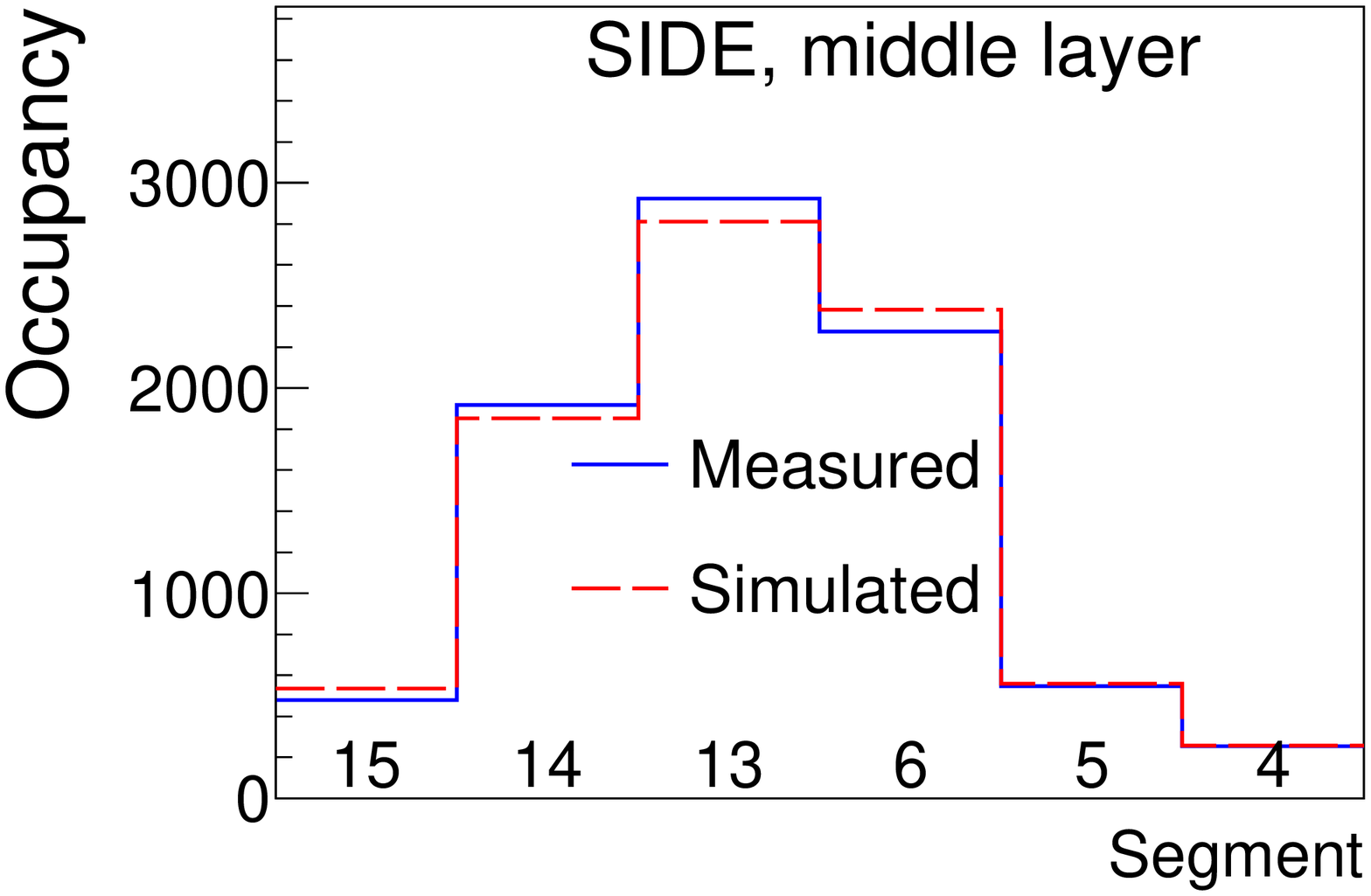}
  \includegraphics[width=0.48\textwidth]{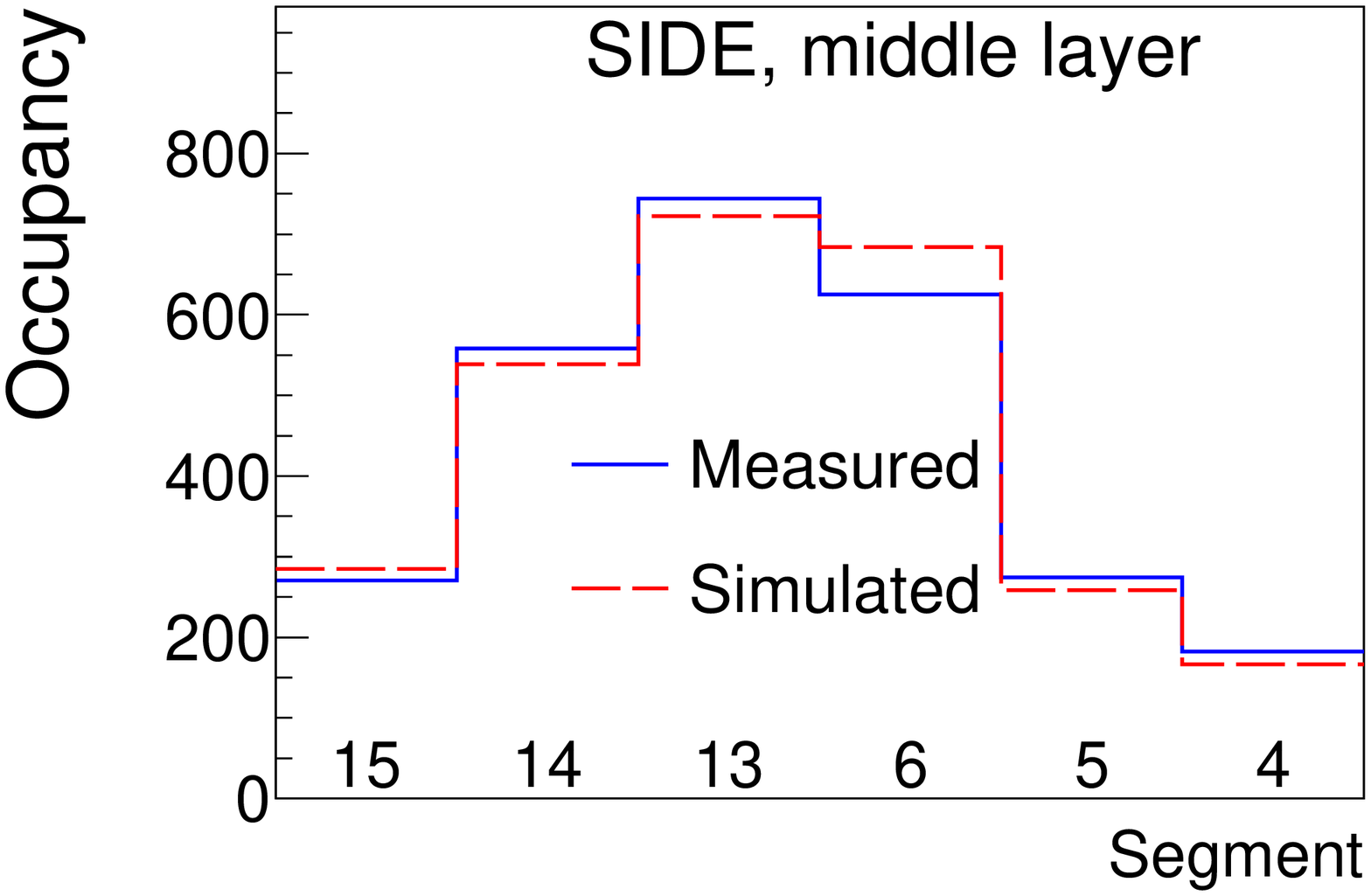}\\

  \includegraphics[width=0.48\textwidth]{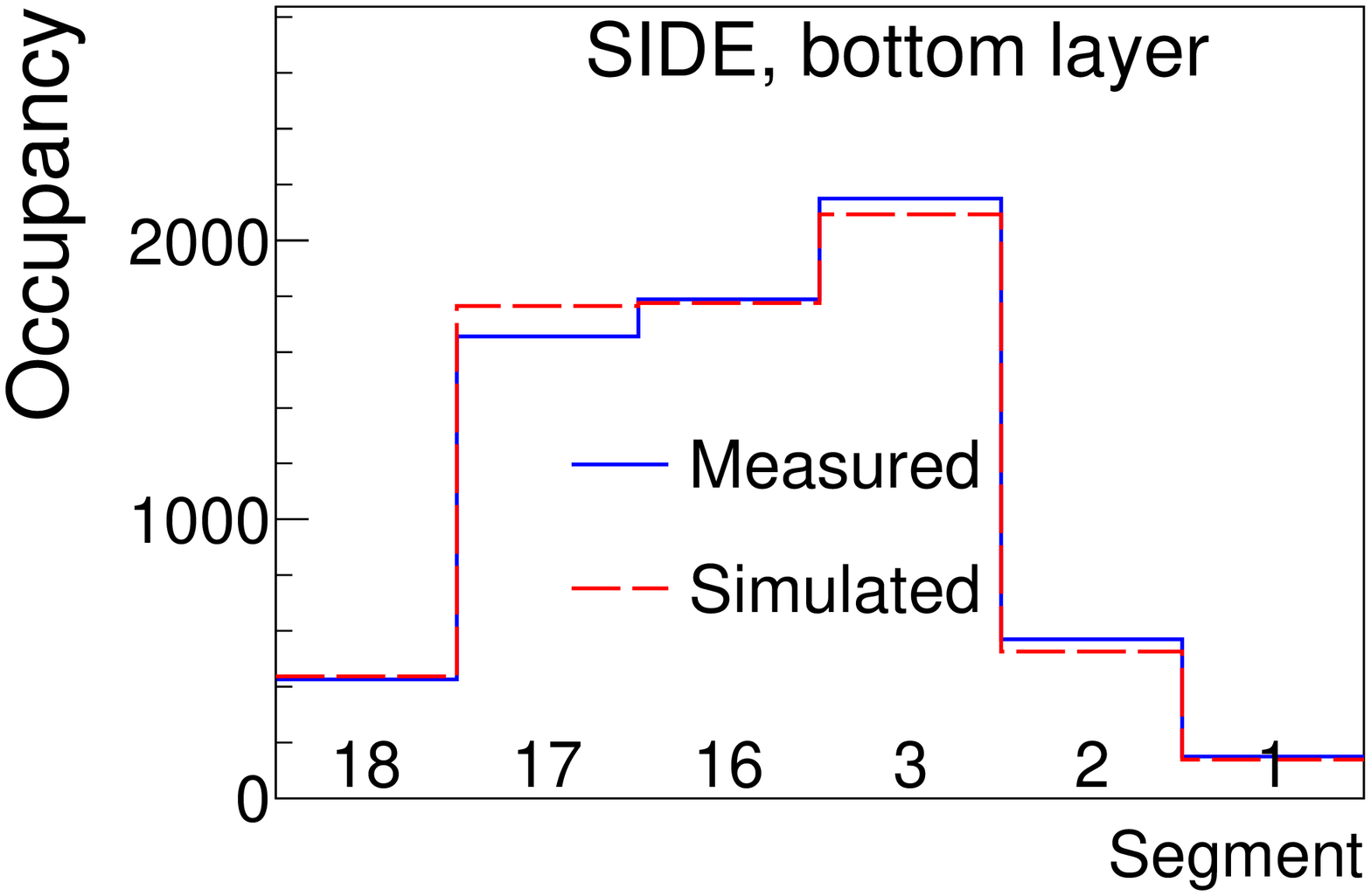}
  \includegraphics[width=0.48\textwidth]{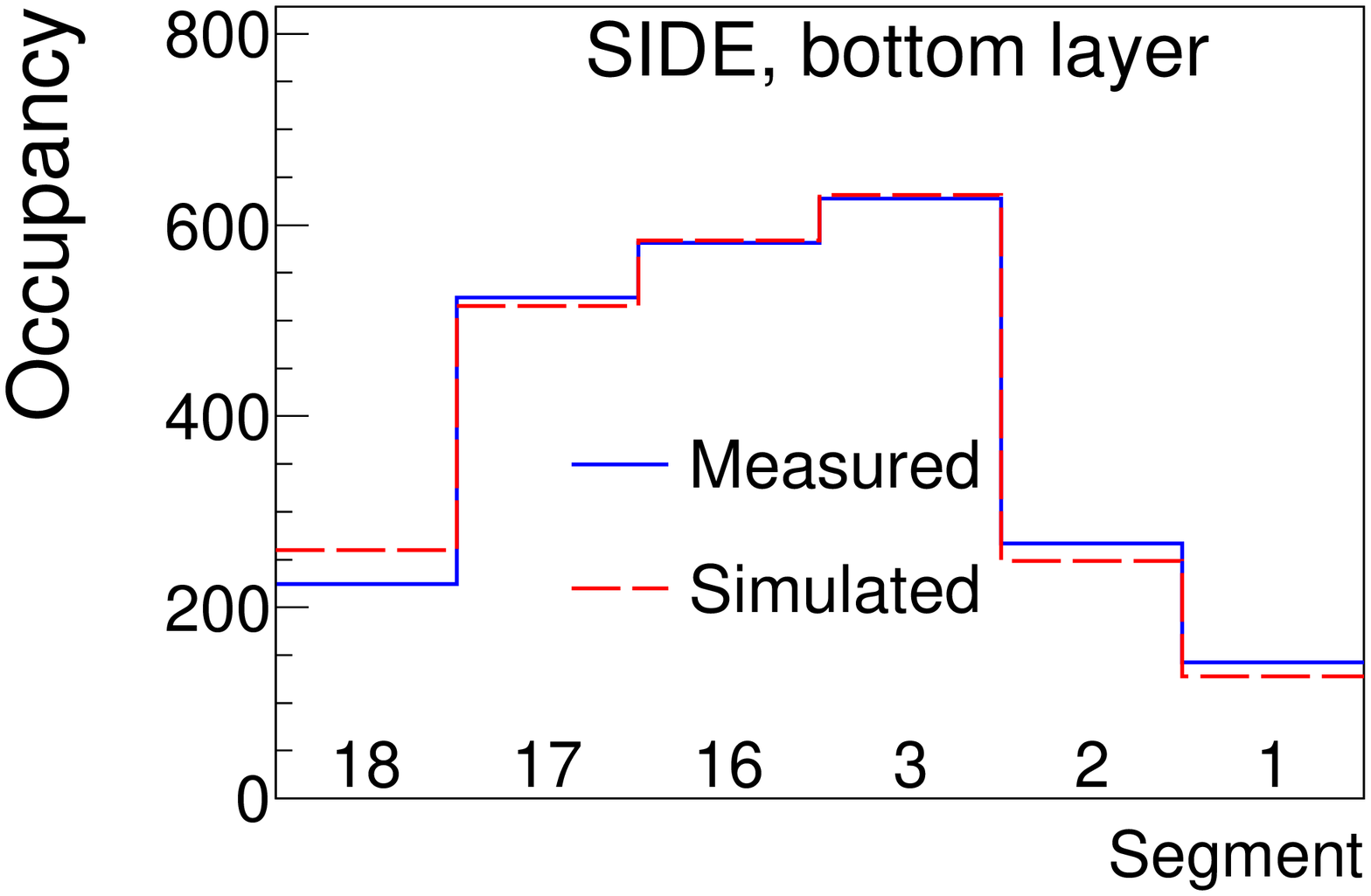}
\flushleft
\hskip 3.6cm (c) \hskip 6.3cm (d)\\
\caption{
Measured occupancies together with the best predictions for
the irradiation with $^{208}$Tl from the side 
for (a) 0.58\,MeV, middle layer
(b) 2.61\,MeV, middle layer
(c) 0.58\,MeV, bottom layer
(d) 2.61\,MeV, bottom layer.
}
\label{fig:occ:res:sTh} 
\end{figure}

This result is barely consistent with the one obtained 
with the scan. 
Figure~\ref{fig:occ:res:sTh} depicts the measured occupancies 
together with the best predictions
for the middle and bottom layer for the $^{208}$Tl irradiation
from the side. 
An irradiation directly from
the side is {\it a priori} not favorable for this method.
However, the patterns are reproduced quite well, 
indicating that probably the misalignment caused by
the remounting of the detector and the changes in the background due to
the movement of the equipment 
were larger than hoped for.
The usage of this data for the analysis presented here was originally
not planned. Therefore the systematic uncertainties 
relevant for this analysis were not really well
controlled. With some care, they certainly could be reduced significantly.

The overall results 
confirm the ability to extract the axes orientation from
the occupancy patterns in a $\phi$-segmented detector, even
if the detector is not in a perfect condition.
The precision depends on the care taken to avoid
systematic uncertainties.
The source position has to be well controlled under all circumstances.
A symmetric configuration reduces the systematic uncertainties.
The background has to be well known, if lines are used that also
occur in the background.
In general, a source should be chosen which has
lines with a reasonably high probability of 
containment in a segment and high enough energy not to be
disturbed too much by the surrounding material. 
The selection will depend on the setup and the detector. 
In the case considered here, the $^{60}$Co lines were
found to be best suited.

\section{Conclusions}

A new method is presented to determine the axes orientation
of $\phi$-segmented germanium detectors. The method relies on the
ability to predict occupancy patterns 
caused by the transverse anisotropy
of the drift of the relevant charge carriers.
The axes orientation of a test detector was 
determined with a precision
of about 6$^\circ$.
The precision was limited by the knowledge about the source
locations. In general, a precision of 5$^\circ$ should be attainable.
This is compatible to the precision achieved with the well
known scan method and is the precision needed for the
usages of pulse shapes as currently envisioned.
The new method can be used
in any setup, also with multiple detectors, using any calibration
source as long as the setup
can be simulated with sufficient accuracy 
and background, if present, is properly taken into account.
Therefore, it becomes unnecessary to perform time consuming scans
to characterize detectors before they are integrated in a
larger device.
The method was developed and tested for true-coaxial detectors.
However, it can be applied to any detector with an appropriate
segmentation scheme. The modifications
for a $\phi$-segmented closed-end cylindrical detector will be minimal.

The method has systematic uncertainties depending on the
precision of the simulation of the setup and of the transverse
anisotropy in the detector crystal.
The dependence of the patterns on the transverse anisotropy
provides the potential
to determine charge carrier mobilities.

\section{Acknowledgements}
% \begin{acknowledgement}

We would like to thank the members of the \textsc{MaGe} Monte
Carlo group for their kind support.

% \end{acknowledgement}

%--------------------------------------------------------
% bibliography
%--------------------------------------------------------
%\clearpage
%\newpage

\end{document}